\documentclass[preprint]{aastex631}
\usepackage[mathlines]{lineno}
\usepackage{amsmath}
\usepackage{url} 
\usepackage{lineno}
\usepackage{graphicx}
\usepackage{color}
\usepackage{physics}
\usepackage{xcolor}
\usepackage{rotating}

\shorttitle{Spatially highly resolved magnetic configuration on Venus}
\shortauthors{He et al.}
\graphicspath{{./}{figures/}}

\begin{document}

 \title{Spatially highly resolved solar-wind-induced magnetic field on Venus} 
 \correspondingauthor{Maosheng He}\email{he@iap-kborn.de}
 \author[0000-0001-6112-2499]{Maosheng He}
 \altaffiliation{Now at Leibniz-Institute of Atmospheric Physics at the Rostock University, Kuehlungsborn, Germany}
 \affiliation{Department of Physics and Earth Sciences, Jacobs University Bremen, Bremen, Germany}

 \author[0000-0001-7843-9214]{Joachim Vogt}
 \affiliation{Department of Physics and Earth Sciences, Jacobs University Bremen, Bremen, Germany}

 \author[0000-0001-8619-187X]{Eduard Dubinin} 
 \affiliation{Max Planck Institute for Solar System Research, Göttingen, Germany}
 \author[0000-0002-0980-6292]{Tielong Zhang} 
 \affiliation{Space Research Institute, Austrian Academy of Sciences, Graz, Austria}

 \author[0000-0003-4609-4519]{Zhaojin Rong} 
 \affiliation{Key Laboratory of Earth and Planetary Physics, Institute of Geology and Geophysics, Chinese Academy of Sciences, Beijing, China}



 \begin{abstract}
The current work investigates the Venusian solar-wind-induced magnetosphere at a high spatial resolution using all Venus Express (VEX) magnetic observations through an unbiased statistical method. We first evaluate the predictability of the interplanetary magnetic field (IMF) during solar’s magnetospheric transits and then map the induced field in a cylindrical coordinate system under different IMF conditions. Our mapping resolves structures on various scales, ranging from the ionopause to the classical IMF drapping. 
We also resolve two recently-reported structures, a low-ionospheric magnetization over the terminator and a global "looping" structure in the near magnetotail. In contrast to the reported IMF-independent cylindrical magnetic field of both structures, our results illustrate their IMF dependence. In both structures, the cylindrical magnetic component is intenser in the hemisphere with an upward solar wind electric field ($E^{SW}$) than in the opposite hemisphere. Under downward $E^{SW}$, the "looping" structure even breaks, which is attributable to an additional draped magnetic field structure wrapping toward $-E^{SW}$. In addition, our results suggest that these two structures are spatially separate. The low-ionospheric magnetization occurs in a very narrow region, at about  88--95$^\circ$ solar zenith angle and 185--210~km altitude. A least-square fit reveals that this structure is attributable to an antisunward line current with 191.1~A intensity at 179$\pm$10~km altitude, developed potentially in a Cowling channel.


 \end{abstract}



 \section{Introduction} \label{sec:intro}
 \begin{figure}
 \centering
 \includegraphics[width=40pc]{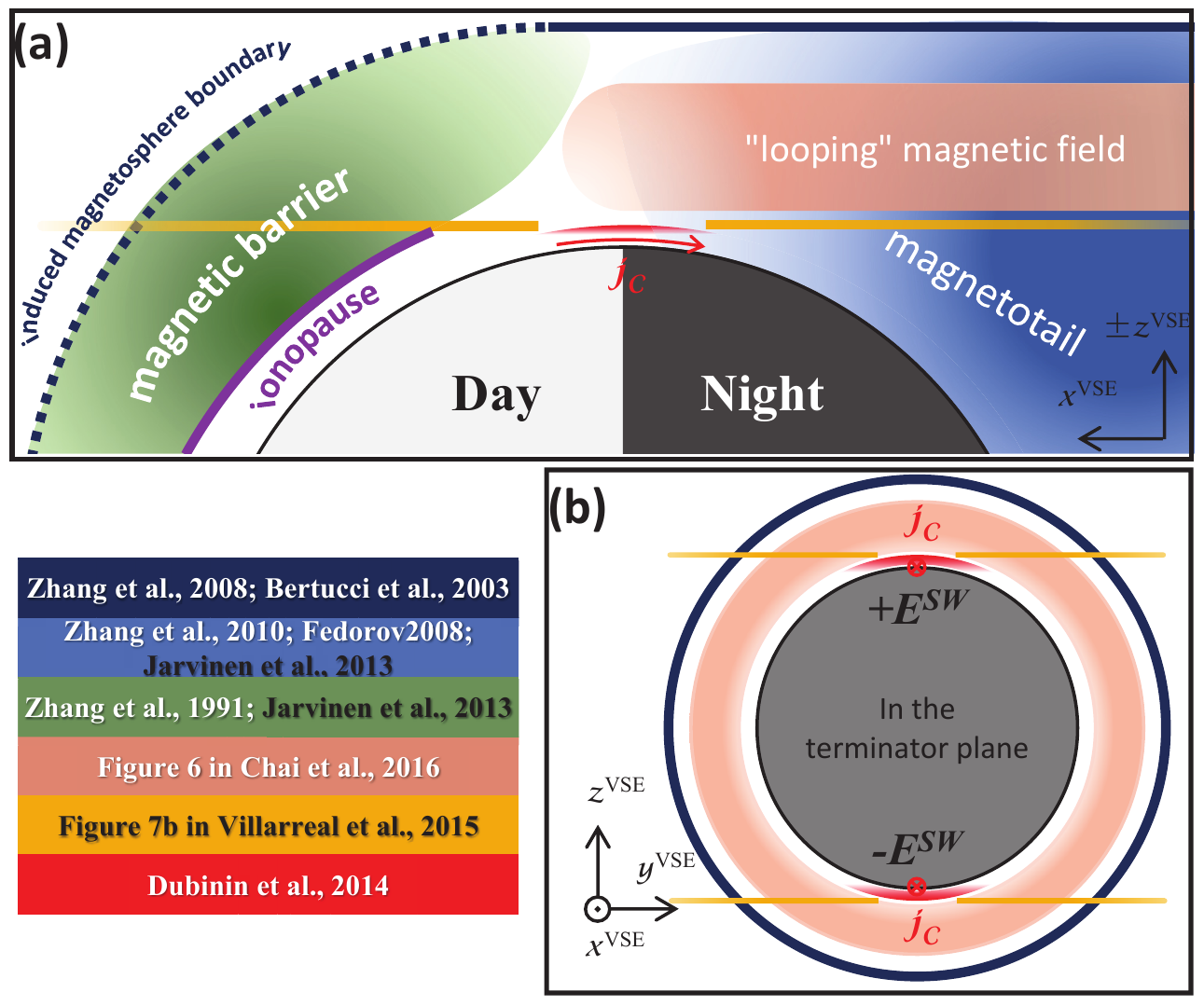}
 \caption{Venusian induced magnetosphere in the planes (a) y=0 (b) and x=0 in the VSE coordinate system. 
  The colored text boxes in the bottom left corner present references for the structures in corresponding colors, in which the white and black text denote observational studies and simulations, respectively. Note that the definition of the dayside induced magnetosphere boundary is denoted here as a dotted line as the relevant definitions are different in different studies. Read Section~\ref{sec:intro} for details. }
  \label{fig:Symtr_sketch}
 \end{figure}

 \begin{figure}[h]
  \centering
  \includegraphics[width=37pc]{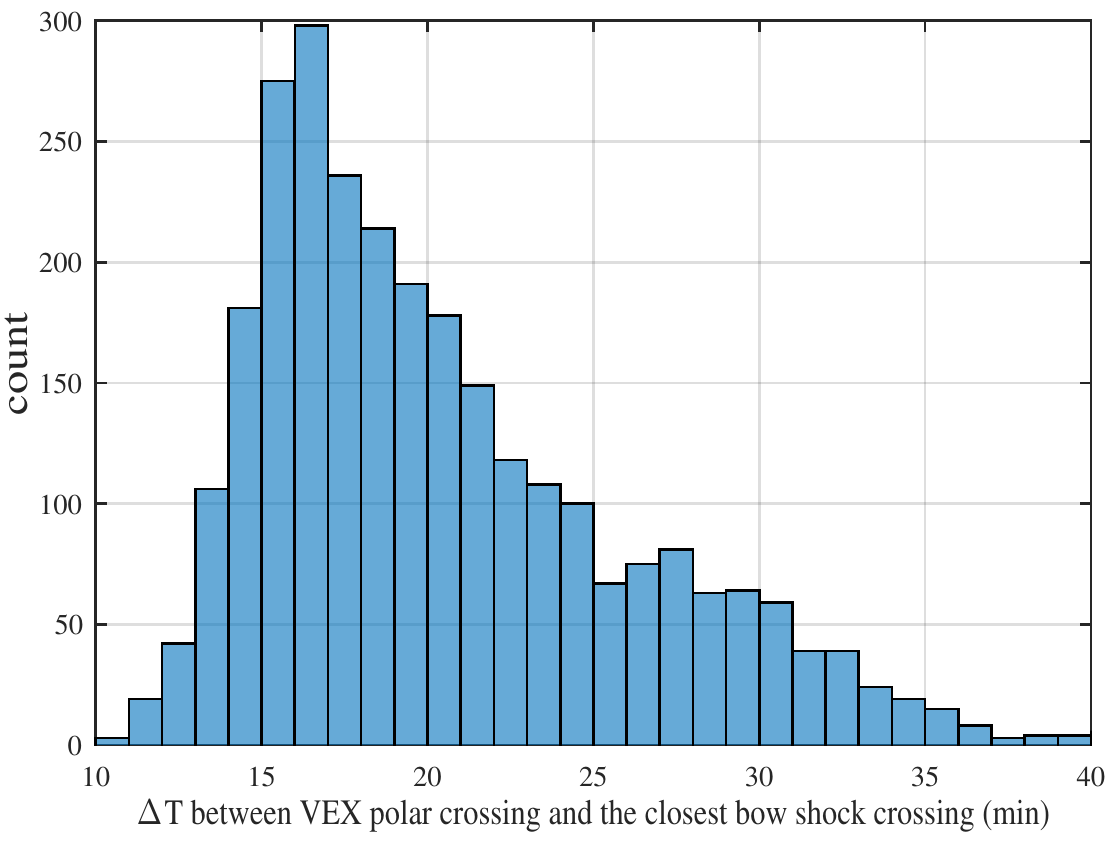}
  \caption{Histogram of the temporal separation between the VEX's maximum latitude passing and the corresponding closest crossing of the bow shock's outer edge. Read Section \ref{sec:IMF} for details.}
  \label{fig:DT_Dist}
 \end{figure}

Some unmagnetized or weakly magnetized bodies in the solar system possess electrically conductive ionospheres, e.g., the planets Mars and Venus and the satellite Titan. These bodies interact with the solar wind (SW), inducing electric currents and shaping comet-tail-like magnetospheres \citep[][]{Bertucci2011}. Among these bodies, Venus does not have Mars-like crustal fields, is relatively close to the sun and surrounded by dense SW, and has been observed using artificial orbiters. Therefore, the Venusian magnetosphere is controlled sensitively by the SW, and relevant observations are relatively abundant, making the Venusian magnetosphere a natural laboratory to study the interactions between the SW and unmagnetized bodies.

The interactions between SW and Venus plasma environment have been investigated observationally \citep[e.g.,][]{Russell2006,Bertucci2011}, and essential processes and structures have been reproduced in global hybrid simulations \citep[e.g.,][]{Jarvinen2013,Jarvinen2014}. Here, we briefly summarize the processes and structures, and for details, readers are referred to the relevant reviews \citep[e.g.,][]{Brace1991,Baumjohann2010,Bertucci2011,Russell2007,Dubinin2020}. The SW plasma surrounding Venus is supersonic and super-Alfvenic (the sound speed and Alfven speed are significantly lower than the solar wind bulk velocity) with the frozen-in interplanetary magnetic field (IMF, $\mathbf{B}^{IMF}$). When the SW plasma encounters Venus, the SW motional electric field ($E^{SW}$) drives electric currents in the highly conducting ionosphere and the surrounding plasma. The magnetic fields associated with the currents decelerate the SW flow, forming a bow shock, magnetosheath, and  magnetic barrier or magnetic pileup region. At the bow shock, SW plasma is decelerated to subsonic. The deceleration thermalizes the plasma and excites intense waves and turbulence. The shocked SW plasma remains collisionless (collision frequencies much smaller than gyro frequencies), which populates the magnetosheath but cannot easily penetrate the magnetic barrier and is mostly deflected around the planet. The deflection gives rise to a wake with low-density plasma downstream, forming the magnetotail, where happen reconnection events and particle acceleration and loss. The boundary of the low-density plasma region is known as the induced magnetosphere boundary (IMB) or sometimes magnetopause. The induced currents distribute most intensely on a thin layer, where the ionospheric thermal pressure holds off the magnetic barrier's magnetic pressure. The layer is the bottom boundary of the magnetic barrier and is known as the ionopause. The ionopause currents are closed through currents in the magnetic barrier or on IMB. According to, e.g., \citet{Russell2007} and \citet{Brace1991}, the magnetic barrier is referred to as the most inner part of the magnetosheath, and both border to the dayside ionosphere at the ionopause, whereas according to, e.g., \citet{Zhang2008} and \citet{Baumjohann2010}, the magnetic barrier is defined as an independent region between the magnetosheath and the ionopause. The boundary between the magnetosheath and the magnetic barrier is characterized by a discontinuity of magnetic distortion and magnetosheath wave terminations, which is referred to as the dayside IMB or magnetopause \citep{Zhang2008}. Referring to \citet{Zhang2008}, we sketch the structures involved in the current study in Figure~\ref{fig:Symtr_sketch}.


The magnetic field configuration of the induced magnetosphere is characterized by a draping configuration, elongating along the Sun-Venus axis \citep[e.g.,][]{Luhmann1986a}. In the plane normal to the Sun-Venus axis, the orientation of the induced magnetic field is mainly determined by $E^{SW}$ and the cross-flow IMF component, which is usually described in a coordinate system defined referring to $E^{SW}$, known as the Venus-Sun electric field coordinate system \citep[VSE for short, also called SW velocity and magnetic field coordinate system, e.g.,][]{Dubinin2014,Du2013}. The x-axis of the VSE coordinate system is antiparallel to the upstream SW flow velocity (${\mathbf{v}^{SW}}$), the y-axis is parallel to the IMF cross-flow component, and the z-axis is parallel to $E^{SW}$, therefore also termed as the $E^{SW}$ axis.
In VSE coordinates, the classical induced magnetic field is overall anti-symmetric between the $\pm$y hemispheres, and symmetric between the $\pm$z hemispheres \citep[namely $\pm E^{SW}$ hemisphere, e.g.,][]{Bertucci2011,He2016}.
The $\pm E^{SW}$ symmetry was reported broking in both the magnetic barrier and the magnetotail: the magnetic field is stronger in the $+E^{SW}$ hemisphere than that in the $-E^{SW}$. These features were investigated in statistical studies \citep{Du2013,Bertucci2003,Bertucci2005,Zhang1991,Zhang2008,Zhang2010} and reproduced in simulations \citep{Brecht1990,Brecht1991,Jarvinen2013}, and therefore have been well understood.

The above-mentioned magnetospheric structures are on planetary scales. Recently, two relatively small-scale magnetospheric structures were observed in the low-altitude ionosphere \citep[below 300~km altitude,][]{Zhang2012,Zhang2015} and in the magnetotail \citep{Chai2016}, referred to as the $\pm E^{SW}$ asymmetrical low-ionospheric magnetization \citep[e.g.,][]{Dubinin2014} and "looping" magnetic field \citep{Chai2016}, as indicated by the red and pink symbols in Figure \ref{fig:Symtr_sketch}, respectively.
The low-ionospheric asymmetry is characterized by a preference of induced magnetic field pointing towards +y in both $\pm E^{SW}$ hemispheres, namely parallel to the cross-flow IMF in the +$E^{SW}$ hemisphere but antiparallel in the $- E^{SW}$ hemisphere, which was also described, alternatively, as a dawnward preference over the Venusian north pole \citep{Zhang2015}. Various mechanisms were proposed, including giant flux ropes formed in the magnetotail \citep{Zhang2012}, Hall currents induced regionally in the low ionosphere \citep{Dubinin2014}, and antisunward transports of low-altitude magnetic belts build-up on the dayside \citep{Villarreal2015}. The "looping" magnetic field is characterized by a planetary-scale cylindrical symmetry of cylindrical magnetic field in the near-Venus magnetotail, which was explained in terms of a global induced current system distributed on double cylindrical layers\citep{Chai2016}. Most studies on these two structures are observational, based on either manually selected cases \citep[e.g.,][]{Dubinin2014, Zhang2015} or observations without case selections \citep{Chai2016}. Studies with selected cases resolved the structures at relatively high spatial resolutions but were potentially subject to prior knowledge in the selection, whereas unbiased studies, without any guiding selection, presented the structures at typically low spatial resolutions. Consequently, the spatial distributions of these structures are not determinedly resolved. For example, it is not known whether the low-ionospheric asymmetry overlaps with the "looping" structure.
To deal with this problem, we present unbiased statistical results of the near-Venus magnetic field at a high spatial resolution up to about 50~km. Our results suggest that, as sketched in Figure \ref{fig:Symtr_sketch}, the low-ionospheric asymmetry was not covered by a recent relevant magnetohydrodynamic simulation in \citet{Villarreal2015}, and the low-ionospheric asymmetry is not overlapped with the "looping" field. We further compare the spatially highly resolved depictions under different IMF conditions, demonstrating that the "looping" structure is not cylindrically symmetric and breaks under some IMF conditions.

Below, after explaining our data analysis in Section \ref{sec:Data analysis}, we demonstrate the capability of our method in resolving small-scale magnetospheric structures in Section \ref{sec:draping} and investigate the "looping" field and low-ionospheric asymmetry in Section \ref{sec:New_lights}.

 \begin{figure}[h]
  \centering
  \includegraphics[width=40pc]{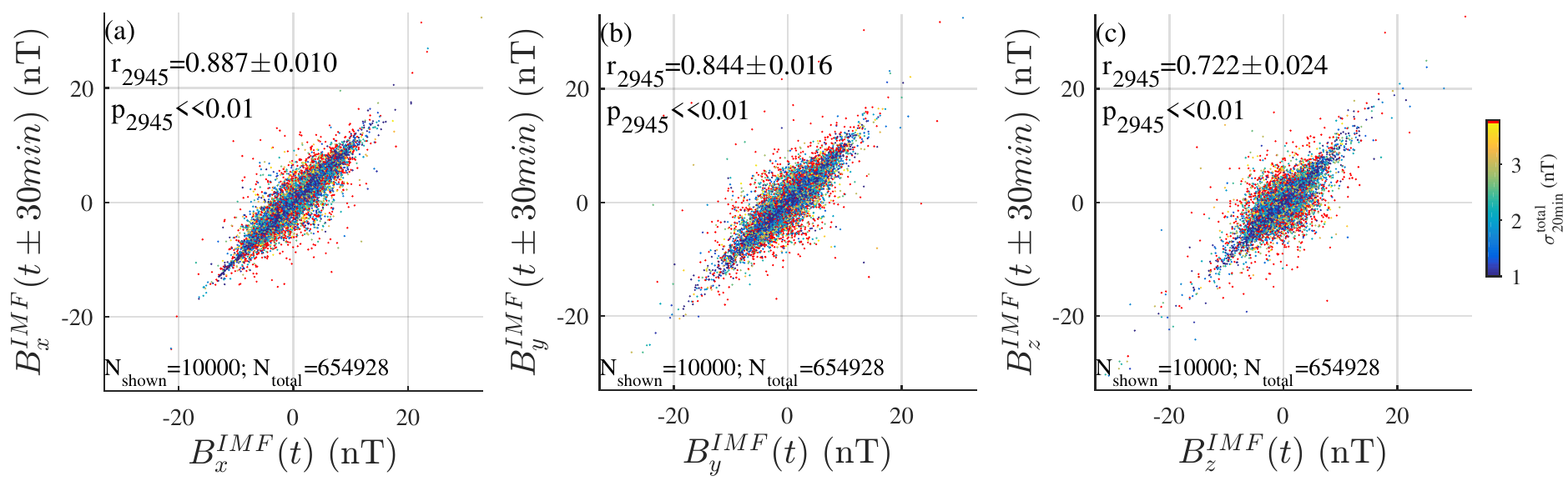}
  \caption{(a) The correlation between 20-min averaged $B_x^{IMF}$ at time $t$ and that at $t\pm$30~min at Venus. The color code represents the IMF disturbance $\sigma_{20\min}^{total}$ defined in Section~\ref{sec:IMF}. The red points display the cases collected during the 15\% most disturbed states characterized by $\sigma_{20\min}^{total}>$3.7~nT. Indicated in the top left corner are the correlation coefficient $r\pm \Delta r$, and $p$-value, estimated from a bootstrapping analysis detailed in Section~\ref{sec:IMF}. For a better visibility, shown is only a portion of randomly selected pairs. (b and c) Same as Panel (a) but for $B_y^{IMF}$ and $B_z^{IMF}$. Read Section \ref{sec:IMF} for details.}
  \label{fig:CorrIMFxyz}
 \end{figure}
 \section{Data analysis} \label{sec:Data analysis}

 Venus Express (VEX) operated in a polar orbit. The orbit had a periapsis at 170--350~km altitude at about 78$^\circ$N latitude in the Venus Solar Orbital coordinate system \citep[VSO, the x-axis points from Venus towards the sun, the z-axis is normal to the Venus orbital plane, pointing northward, and the y-axis completes the right-handed set, e.g.,][]{Svedhem2007, Titov2006}. Due to the high orbital eccentricity, VEX repeatedly encountered SW, magnetosheath, and magnetosphere. VEX was not accompanied by any independent spacecraft for monitoring the IMF condition in VEX's magnetospheric transit. Therefore, the IMF has to be estimated. Here, after introducing and evaluating our IMF prediction in Section \ref{sec:IMF}, we introduce a cylindrical coordinate system, our IMF binning, and a high spatial resolution mapping approach in Sections \ref{sec:Coord}, \ref{sec:IMFconditions}, and \ref{sec:Mapping}, respectively. We use all available 4-s-averaged magnetic field observations from about 3000 orbits during VEX's eight Venusian years of operation \citep{Zhang2006a}.
 Readers are referred to \citet{He2016} for the distribution of the orbits as functions of local time, solar zenith angle, and altitude.

 \subsection{IMF predictability for VEX's magnetospheric transit} \label{sec:IMF}

 A popular IMF estimation approach compares the IMF conditions immediately before the inbound bow shock crossing and after the outbound crossing \citep[e.g.,][]{Zhang2015}.
 However, the current study focuses on the magnetosphere over the northern polar cap, which is typically much closer to one bow shock crossing, either inbound or outbound, than the other in any given orbit. This situation is similar to that illustrated by \citet{He2017} where the authors estimated IMF for MESSENGER's Mercurian magnetospheric transit using the actual IMF observation from the nearest SW encounter. Following the IMF estimation approach in \citet{He2017}, we estimate the IMF conditions as the average of actual IMF observations in a remote 20-min-wide interval of VEX's nearest SW encounter at the closest bow shock crossing, either inbound or outbound.  The bow shock crossings are identified manually and detailed in \citet{Rong2014}.
 In each orbit, we first identify the VEX's maximum latitude arriving as a reference point of the magnetospheric transit, and calculate the temporal separation between the reference point and the corresponding closest bow shock crossing. The temporal separations of all orbits are distributed in Figure~\ref{fig:DT_Dist}, which peaks at about 17~min with an average of 20.5~min. By average, the reference point is separated away from the center of the SW interval by $\Delta t$= $\delta t$/2+20.5~min =30.5~min, where $\delta t$=20~min is the sampling window width. 
 Approximately IMF is estimated as the actual IMF observations 30~min preceding or succeeding. Below, we evaluate the estimation using actual IMF observations through a correlation analysis.

 For the correlation analysis, we first select the magnetic field measured in the SW and average them within discrete 20-min-wide intervals.
We denote $\mathbf{B}^{IMF}(t)$ as the average in the interval from $t-10$min to $t+10$min. Then, $\mathbf{B}^{IMF}(t\pm30$min) are 'estimations' of $\mathbf{B}^{IMF}(t)$. In total, the VEX SW observations enable about $N_{pair}$=6.5$\times$ 10$^5$ independent pairs of [$\mathbf{B}^{IMF}(t)$, $\mathbf{B}^{IMF}(t\pm30$min)], the three components of which are scatter-plotted in the three Panels in Figure~\ref{fig:CorrIMFxyz}. The points are color-coded according to the IMF variability defined as $\sigma _{20\min}^{total}$
 $=\sqrt{\sigma _{Bx}^2+\sigma _{By}^2+\sigma _{Bz}^2}$.
Here $\sigma _{Bx}$, $\sigma _{By}$, and $\sigma _{Bz}$ are the standard deviation of the three magnetic components within the 20-min sampling window centering at $t\pm30$min. According to ascending $\sigma _{20\min}^{total}$ and following \citet{He2017}, we sort the $N_{pair}$=6.5$\times$ 10$^5$  pairs into ten groups evenly, and calculate the correlation coefficient $r$ between $\mathbf{B}^{IMF}(t)$ and $\mathbf{B}^{IMF}(t\pm30$min) in each group, illustrating that $r$ decreases with increasing $\sigma _{20\min}^{total}$ (not shown). For a robust analysis \citep[following Section 2.5,][]{He2017}, we exclude the VEX orbits associated with disturbed IMF states, characterized by $\sigma _{20\min}^{total}>$3.7~nT, from the following analyses. In the end, we get $N_o$=2945 orbits for which bow shock crossings are identified when IMF states are not disturbed.

 After removing the 15\% most disturbed IMF pairs from the $N_{pair}$=6.5$\times$ 10$^5$ pairs, we carry out a bootstrapping analysis as follows.
 First, we select randomly $N_o$=2945 samples from the reduced set of 6.5$\times$10$^5$ pairs with replacement to calculate the correlation coefficient $r$ and $p$-value.
 The sampling is repeated for $N_s$=5000 times, yielding $N_s$ values of $r$ and $p$. Here, $N_s$=5000$>N_o$ is selected subjectively. 
 The average and standard deviation of $r$ are present in each panel of Figure~\ref{fig:CorrIMFxyz}, indicating the correlations are significant for all IMF components and demonstrating that the IMF estimation is reasonable. The correlation coefficients of the three components are close to each other, suggesting that their predictability is comparable. This is different from the IMF predictability at Mercury \citep{He2017} where different components' predictability is significantly different.

 \subsection{A cylindrical coordinate system for cataloging IMF conditions}
 \label{sec:Coord}
 Venus does not have a significant intrinsic magnetic field, neither global dipole field nor regional crustal field, which is magnetically spherically symmetrical.
 The interaction between SW and Venusian ionosphere induces a global magnetic field $\mathbf{B}^V$ and a magnetosphere. The interaction is largely magnetically rotationally symmetrical with respect to the axis along ${\mathbf{v}^{SW}}$. Below, we describe the symmetry mathematically and use it to discuss the VSE system and introduce a new coordinate system.

 \begin{figure}[h]
\centering
\includegraphics[width=40pc]{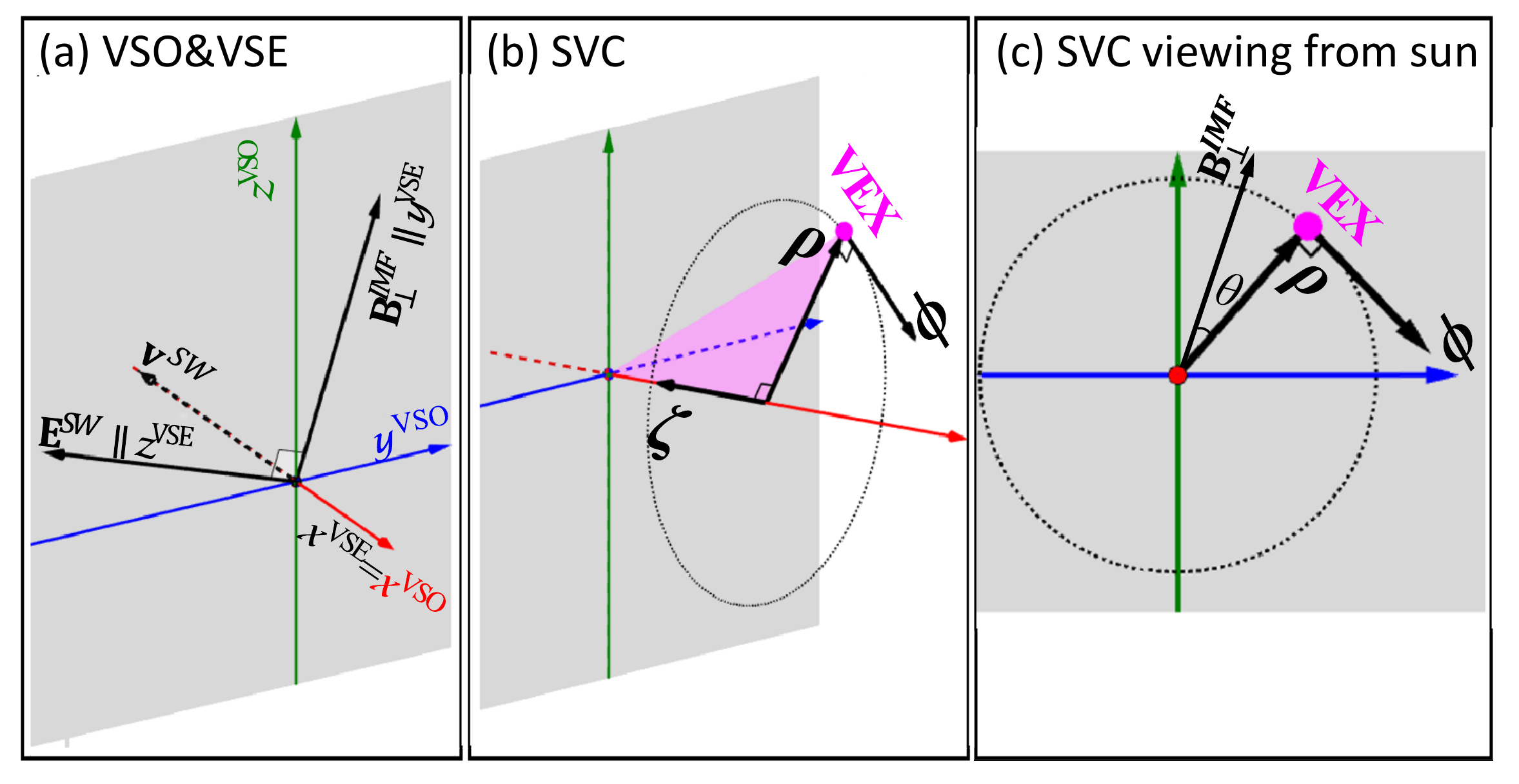}
\caption{
 (a) VSO (colored axis) and VSE (black axis) coordinate systems.
 (b) VSO (colored axis) and Solar-Venus-cylindrical (SVC, black axis) coordinate systems.
 (c) same as (b) but viewing from the sun.
 In each panel, the red axis points from the Venusian center to the sun while the green axis is normal to the Venus orbital plane and points to the north. Read Section \ref{sec:Coord} for details.} \label{fig:Coord}
 \end{figure}

\begin{figure}[h]
 \centering
 \includegraphics[width=38pc]{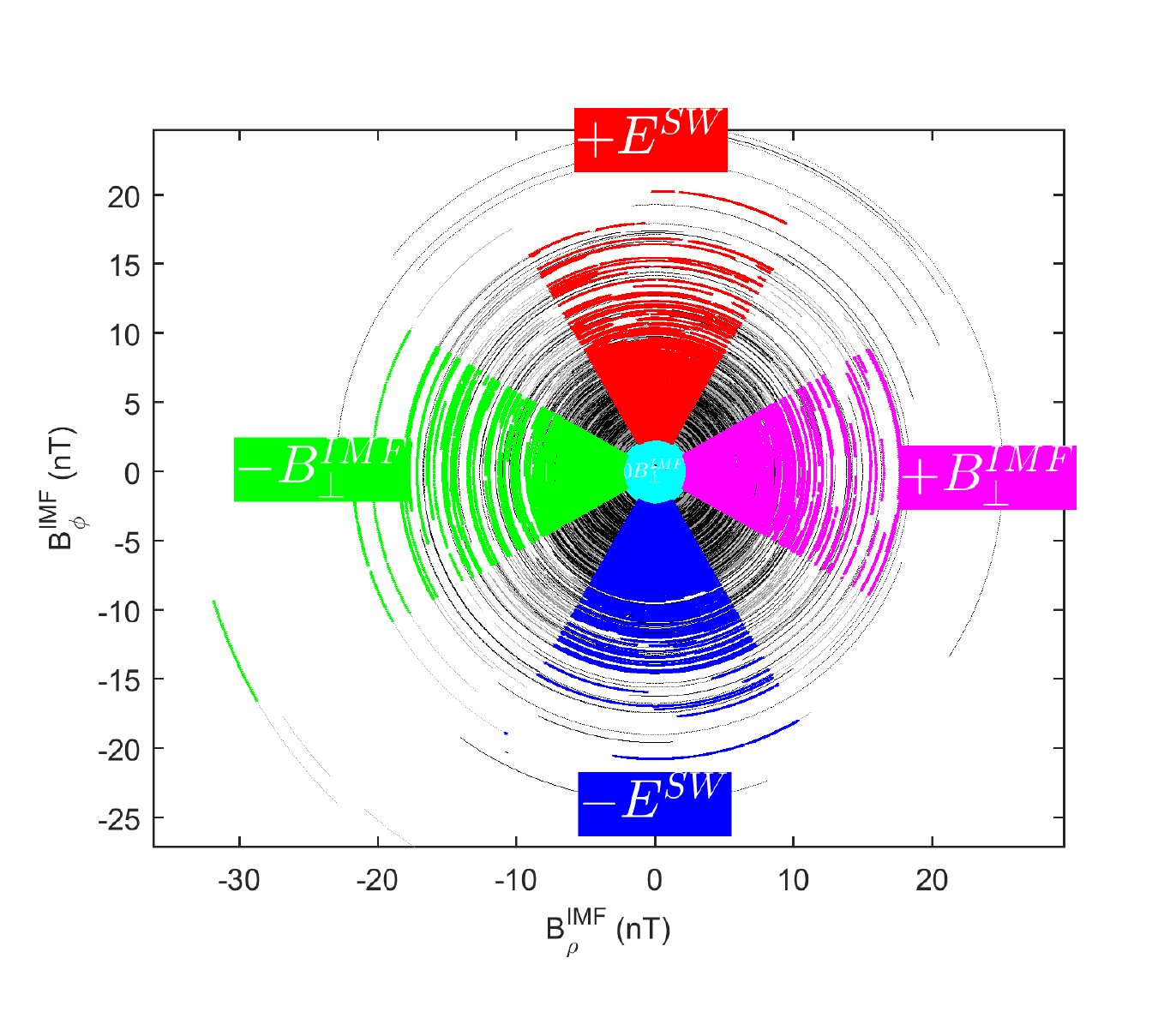}
 \caption{ Estimations of the cross-flow IMF component for all the VEX's magnetospheric transits in the Solar-Venus-cylindrical (SVC) coordinate system. 
 Each point corresponds to one 4-s-resolved magnetic vector collected by VEX under 4400~km altitude and within the solar zenith angle between 65$^\circ$ and 150$^\circ$, while each arc corresponds to a transit of the targeting region of one orbit. The five colors represent five SW/IMF conditions, characterized by different polarities of $\mathbf{B}_\perp^{IMF}$ and $\mathbf{E}^{SW}$. Read Section \ref{sec:IMFconditions} for details.
 }
 \label{fig:IMF_Dist}
\end{figure}

\begin{figure}[h]
 \centering
 \includegraphics[width=40pc]{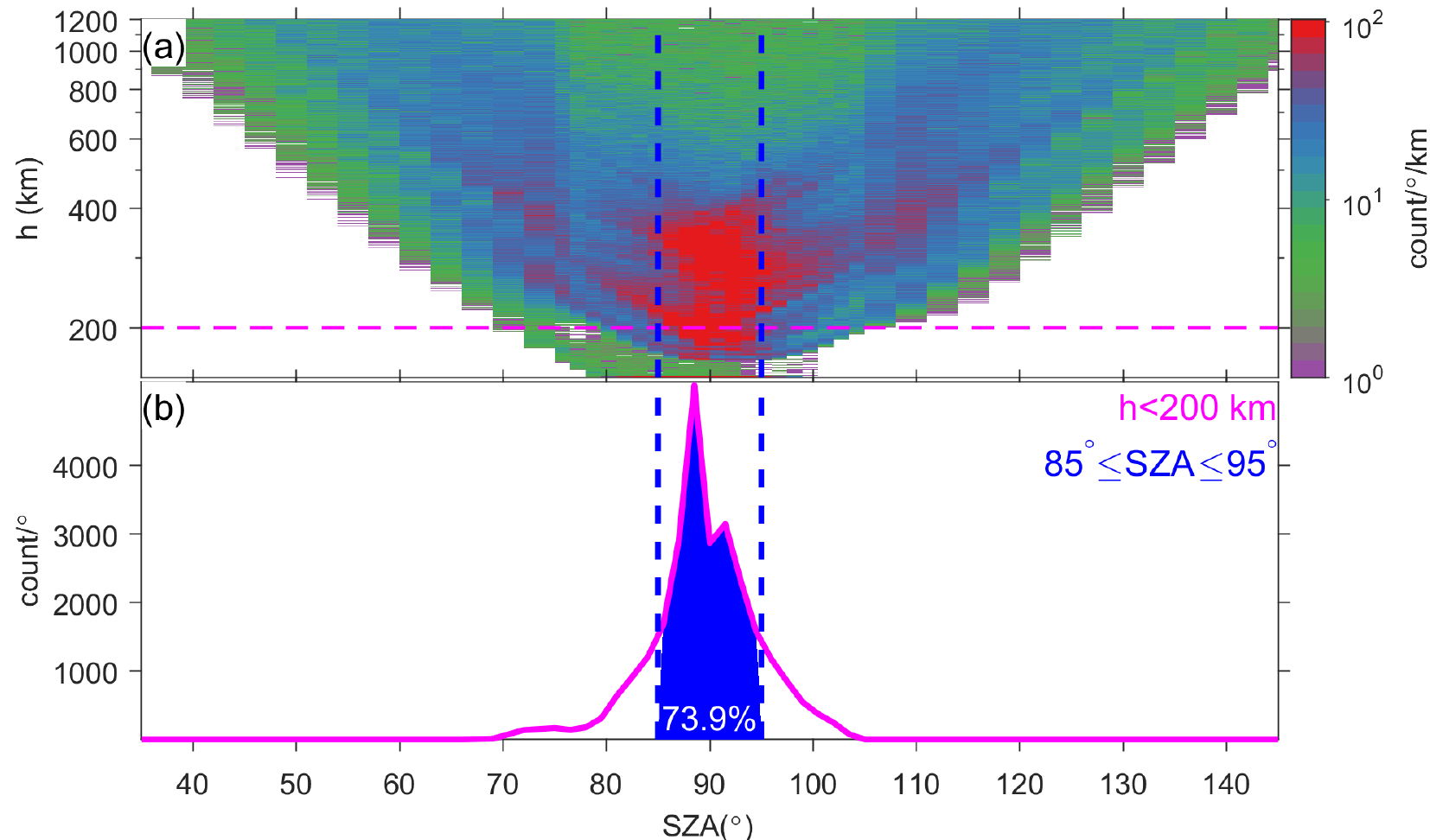}
 \caption{ Histogram of (a) VEX 4-s-resolved magnetic vectors in SZA-altitude depiction and (b) those below 200~km.
 }
 \label{fig:histo}
\end{figure}

 \subsubsection{The rotational symmetry in VSO}
 The large-scale topology of $\mathbf{B}^V$ is characterized by a draped configuration, determined mainly by the upstream SW conditions, mainly its magnetic field $\mathbf{B}^{IMF}$ \citep[e.g.,][]{He2016}. Mathematically,
 \begin{equation}\label{Eq0}
 \mathbf{B}^V=\mathbf{B}^V\left(\mathbf{r}| \mathbf{B}^{IMF}\right)
 \end{equation}
 Here, $\mathbf{B}^{IMF}$ is assumed to be homogeneous on the planetary scale at Venus. Therefore, it is not a function of $\mathbf{r}$ but a function of time $\mathbf{B}^{IMF}\left(t\right)$.

 Investigating the details of Equation \ref{Eq0} is a sort of an essential task of SW and Venus interactions.
 The independent variables of Equation \ref{Eq0} are two vectors, which could be denoted with six scale variables: the position vector $\mathbf{r}:=(r_x, r_y, r_z)$ and $\mathbf{B}^{IMF}:=(B^{IMF}_x,B^{IMF}_y,B^{IMF}_z)$.
 In VSO, the six scale variables could be reduced to five, or a three- and a two-dimensional vector, $\mathbf{r}$ and $\mathbf{B}_\perp^{IMF}:=(0,B^{IMF}_y,B^{IMF}_z)$,  if we neglect $B^{IMF}_x$ for a simplification. (Although $B^{IMF}_x$ also plays an important role in shaping the magnetosphere \citep{Zhang2009}, it is of less capability in explaining the induced magnetic variance near Venus \citep[as quantified in Figure 10 in][]{He2016}, and therefore is often neglected for a first-order approximation.) Accordingly, Equation \ref{Eq0} is often simplified as,
 \begin{equation}\label{Eq1}
 \mathbf{B}^V=\mathbf{B}^V\left(\mathbf{r}| \mathbf{B}_\perp^{IMF}\right)
 \end{equation}
 $\mathbf{B}^V$'s response to the IMF is characterized by rotational symmetry. Mathematically, for any angle $\omega$,
 \begin{equation}\label{RotaSym}
 \mathbf{R}\left(\omega \right)\mathbf{B}^V\left(\mathbf{r}| \mathbf{B}_\perp^{IMF}\right)=\mathbf{B}^V\left(\mathbf{R}\left(\omega \right)\mathbf{r}| \mathbf{R}\left(\omega \right)\mathbf{B}_\perp^{IMF}\right)
 \end{equation}
Here, a product of a vector with the rotation matrix $\mathbf{R}\left(\omega \right):=\left[ \begin{matrix}
 1 & 0 & 0 \\
 0 & \cos \omega & -\sin \omega \\
 0 & \sin \omega & \cos \omega \\
 \end{matrix} \right]$ represents a rotation of the vector with respect to the x-axis (i.e., $-\mathbf{v}^{SW}$) by $\omega$.

 \subsubsection{In VSE}
 Specially, we define the IMF clock angle $\theta^{IMF}:= \arctan(B^{IMF}_y,B^{IMF}_z)$.
 Substituting $-\theta^{IMF}$ for $\omega$ in Equation \ref{RotaSym} yields,
 \begin{equation}\label{RotaSymVSE1}
 \mathbf{R}\left(-\theta^{IMF} \right)\mathbf{B}^V\left(\mathbf{r}| \mathbf{B}_\perp^{IMF}\right)=\mathbf{B}^V\left(\mathbf{R}\left(-\theta^{IMF} \right)\mathbf{r}| \mathbf{R}\left(-\theta^{IMF} \right)\mathbf{B}_\perp^{IMF}\right)
 \end{equation}
 For an arbitrary vector $\mathbf{a}$, its product with $\mathbf{R}\left(-\theta^{IMF}\right)$ defines a coordinate transformation from VSO to VSE: $\mathbf{a}^{VSE}={\mathbf{R}\left(-\theta^{IMF}\right){\mathbf{a}}^{VSO}}$ (see Figure \ref{fig:Coord}a). Therefore, Equation \ref{RotaSymVSE1} writes $\left( \mathbf{B}^V\right)^{VSE}=\mathbf{B}^V\left(\mathbf{r}^{VSE}| \left(\mathbf{B}_\perp^{IMF}\right)^{VSE}\right)$, which could be simplified as
 \begin{equation}\label{RotaSymVSE2}
 \left( \mathbf{B}^V\right)^{VSE}=\mathbf{B}^V\left(\mathbf{r}^{VSE}|\|\mathbf{B}_\perp^{IMF}\|\right)
 \end{equation}
 since $\left(\mathbf{B}_\perp^{IMF}\right)^{VSE} \equiv(0,\|\mathbf{B}_\perp^{IMF}\|,0)$. %
 Equation \ref{RotaSymVSE2} has four independent scale variables, one less than in the VSO representation in Equation \ref{Eq1}, which can therefore facilitate data processing, analysis and interpretation. However, the VSE position vector $\mathbf{r}^{VSE}$ is determined using the IMF estimation. Therefore, the $\mathbf{r}^{VSE}$ determination is subject to the error in the IMF estimation and $\mathbf{r}^{VSE}$ cannot be determined at a singular point $\|\mathbf{B}_\perp^{IMF}\|\approx0$. To conquer these difficulties, in the following subsection, we develop a cylindrical coordinate system.

 \subsubsection{A cylindrical coordinate system}
The previous subsection uses the IMF clock angle $\theta^{IMF}$ to define the rotation matrix $\mathbf{R}\left(-\theta^{IMF}\right)$. We can also define
$\mathbf{R}\left(-\theta^{VEX}\right)$ using
 the clock angle $\theta^{VEX}:= \arctan(r^{VEX}_y,r^{VEX}_z)$ where $\mathbf{r}^{VEX}:=[r^{VEX}_x,r^{VEX}_y,r^{VEX}_z]$ is VEX's position vector. Then, the product $\mathbf{R}\left(-\theta^{VEX}\right)\mathbf{a}^{VSO}$ defines a coordinate transformation of the vector $\mathbf{a}^{VSO}$ from VSO to a cylindrical coordinate system along the VSO x-axis. The three components of $\mathbf{R}\left(-\theta^{VEX}\right)\mathbf{a}^{VSO}$ correspond, in order, to the axial, azimuth, and radial directions, which could be arranged into the conventional order ($\rho$, $\phi$, and $\zeta$) of cylindrical coordinate systems by multiplying with a permutation matrix $\mathbf{U}=: \begin{bmatrix}
 0& 1&0 \\0& 0& -1\\-1& 0& 0 \\ \end{bmatrix}$. As sketched in Figures \ref{fig:Coord}b and \ref{fig:Coord}c, the $\zeta$-axis is the cylindrical axis pointing from the sun to Venus, the $\rho$-axis is the radial direction parallel to the position vector ${{\mathbf{r}}^{VEX}_\perp}:=(0,r^{VEX}_y,r^{VEX}_z)$, and the $\phi$-axis is azimuth coordinate normal to the $\rho$-$\zeta$ plane. This coordinate system is called hereafter as Solar-Venus-cylindrical (SVC) coordinate system.
The product $\mathbf{a}^{SVC}$:=$\mathbf{U}\mathbf{R}\left(-\theta^{VEX}\right)\mathbf{a}^{VSO}$  transforms the vector $\mathbf{a}^{VSO}$ from VSO to SVC.  Specifically, VEX's position in SVC reads [$r^{VEX}_{\rho}, r^{VEX}_{\phi}, r^{VEX}_{\zeta}$]:=
 $\mathbf{U}\mathbf{R}\left(-\theta^{VEX}\right)\mathbf{r}^{VEX}$. Substituting the definitions of $\mathbf{R}$ and $\mathbf{U}$ results in $r^{VEX}_\phi\equiv0$. (Geometrically, $r^{VEX}_\phi\equiv0$ denotes that the VEX is on the plane determined by the sun, Venus, and VEX.) Therefore, all VEX observations distribute two-dimensionally on the $\rho$-$\zeta$ plane in the SVC system.

 Substitute $\mathbf{U}\mathbf{R}\left(-\theta^{VEX}\right)$ for $\mathbf{R\left(\omega\right)}$ in Equation \ref{RotaSym}, yielding, $\mathbf{U}\mathbf{R}\left(-\theta^{VEX}\right)\mathbf{B}^V\left(\mathbf{r}| \mathbf{B}_\perp^{IMF}\right)=\mathbf{B}^V\left(\mathbf{U}\mathbf{R}\left(-\theta^{VEX} \right)\mathbf{r}| \mathbf{U}\mathbf{R}\left(-\theta^{VEX} \right)\mathbf{B}_\perp^{IMF}\right)$
 which could be denoted as
  \begin{equation}\label{RotaSymSVC01}
  \left(\mathbf{B}^V\right)^{SVC}=\mathbf{B}^V\left(\mathbf{r}^{SVC}| \left( \mathbf{B}_\perp^{IMF}\right)^{SVC}\right) =\mathbf{B}^V\left(\mathbf{r}^{SVC}|B_{\rho}^{IMF},B_{\phi}^{IMF}\right)
  \end{equation}
Distributing all on the $\rho$-$\zeta$ plane, the VEX observations allow only investigating the induced field $\mathbf{B}^V$ only on the $\rho$-$\zeta$ plane. Accordingly, Equation \ref{RotaSymSVC01} writes
\begin{equation}\label{RotaSymSVC2}
 \left(\mathbf{B}^V\right)^{SVC}=\mathbf{B}^V\left({r_{\rho}}, {r_{\zeta}}|B_{\rho}^{IMF},B_{\phi}^{IMF}\right)
 \end{equation}

 Similar to Equation \ref{RotaSymVSE2} in VSE, Equation \ref{RotaSymSVC2} also comprises four independent variables and describes the response of $\mathbf{B}^V$ to $B_{\rho}^{IMF}$ and $B_{\phi}^{IMF}$ on the $\rho$-$\zeta$ plane. The SVC position vector is not subject to the IMF estimation and therefore is capable of handling the VSE singular point at $\|\mathbf{B}_\perp^{IMF}\|\approx0$. In the following subsections, we construct $\mathbf{B}^V({r_{\rho}}, {r_{\zeta}})$ under five different $\left( \mathbf{B}_\perp^{IMF}\right)^{SVC}$ conditions.

 \subsection{Five IMF conditions}
 \label{sec:IMFconditions}

 Figure \ref{fig:IMF_Dist} displays the IMF estimations for all VEX magnetospheric transits in SVC $\rho$-$\zeta$ plane. Each point corresponds to one vector of 4-s-resolved $\mathbf{B}^V$ observed below the altitude $h<$4400~km and within the range of solar zenith angle (SZA) between 65$^\circ$ and 150$^\circ$, which is the targeting region of the current work. The arc-shaped traces reflect cylindrical rotations of VEX with respect to $\mathbf{B}_\perp^{IMF}$. One arc represents once targeting region transit. The portion colored in cyan corresponds to the 15\% lowest magnitude $\|\mathbf{B}_\perp^{IMF} \|$, while the rest portions, in magenta, blue, green, and red denote four $60^\circ$-wide ranges of the clock angle $\theta^{IMF}$.
 The five colors represent approximately the five SW/IMF conditions, namely, ($B_{\rho}^{IMF}\approx$0, $B_{\phi}^{IMF}\approx$0), ($B_{\rho}^{IMF}>$0, $B_{\phi}^{IMF}\approx$0), ($B_{\rho}^{IMF}\approx$0, $B_{\phi}^{IMF}<$ 0), ($B_{\rho}^{IMF}<$0, $B_{\phi}^{IMF}\approx$0), and ($B_{\rho}^{IMF}\approx$0, $B_{\phi}^{IMF}>$0). 
 The last four conditions correspond to the transit of the polar caps of $+B_{\perp}^{IMF}$, $-E^{SW}$, $-B_{\perp}^{IMF}$, and $+E^{SW}$ hemispheres in the VSE coordinate system. Accordingly, these SW/IMF conditions are denoted hereafter as $0B_{\perp}^{IMF}$,$+B_{\perp}^{IMF}$, $-E^{SW}$, $-B_{\perp}^{IMF}$, and $+E^{SW}$.

 Under each of these conditions, we combine the VEX observations to construct $\mathbf{B}^V$ in $\rho$-$\zeta$ depiction, using the method detailed in the following subsection.

 \subsection{An non-uniform high spatial resolution method for mapping $\mathbf{B}^V(\rho,\zeta)$}
 \label{sec:Mapping}
\begin{sidewaysfigure}
\centering
\includegraphics[width=23cm]{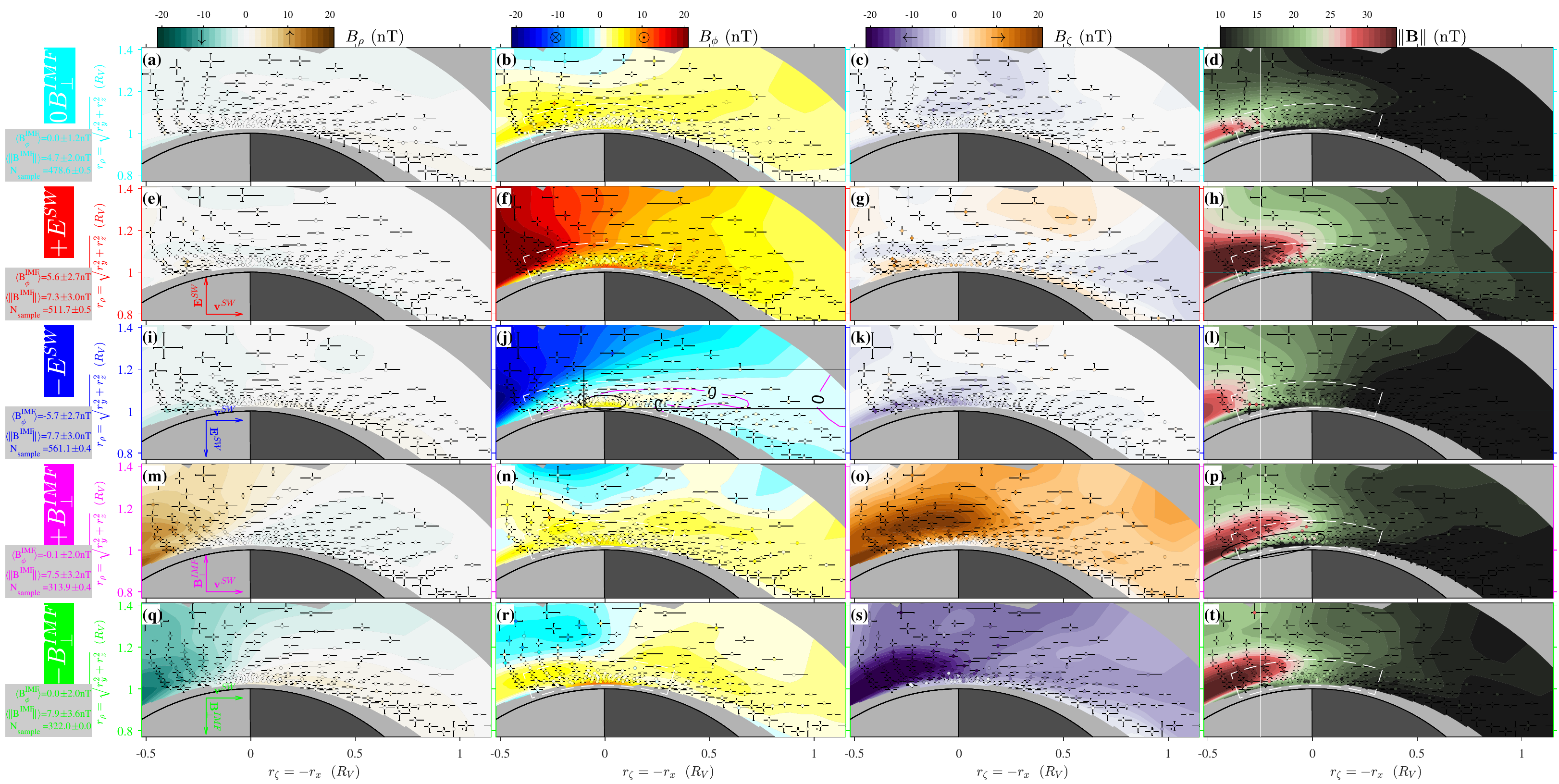}
\caption{ Solar wind (SW) induced magnetic field $\mathbf{B}^V$ in the sun-Venus-VEX plane (namely $\rho$-$\zeta$ plane in SVC ) under the five SW/IMF conditions defined in Figure \ref{fig:IMF_Dist}. One SW/IMF condition is arranged in Panels in one row, as specified in the colored boxes on the very left of each row. Below each box, $\langle B_{\phi}^{IMF}\rangle$ and $\langle\|\mathbf{B}_\perp^{IMF} \|\rangle$ specify the average $\phi$-component and magnitude of the corresponding IMF (namely, the IMF samplings in the corresponding color in Figure \ref{fig:IMF_Dist}), while $N_{sample}$ counts the number of the samplings in one data bin denoted as one black cross in the row. From left to right, the columns display in order the three magnetic components and the magnitude, namely, $B_{\rho}^V$, $B_{\phi}^V$, $B_{\zeta}^V$, and $\|\mathbf{B}^V\|$, respectively. The arrows in (e,i,m,q) represent the orientations of $\mathbf{B}_\perp^{IMF}$ and $\mathbf{E}^{SW}$.
In each panel, the solid black lines display the solid surface of Venus and the terminator; one black cross represents one sampling bin, consisting of vertical and horizontal error bars. The tips and intersecting point of the error bars denote the quartiles of $r_{\rho}$ and $r_{\zeta}$ of the corresponding VEX samplings within the bin. On the intersecting point, the open circle is color-coded according to the median of the corresponding magnetic component within the bin. In the color-code bar on top of Panels (a)--(c), the arrows and circled-cross and -point symbols indicate the direction of the corresponding magnetic field with respect to the $\rho$-$\zeta$ plane. In (h) and (l), the cyan line indicates the region covered by the simulation in Figure 7b by \citet[][]{Villarreal2015}, on top of which the black dashed line around the terminator represents the region excluded in the simulation. Read Section \ref{sec:Mapping} for details.
}
\label{fig:Bxyz5IMF}
 \end{sidewaysfigure}

\begin{figure}[h]
\includegraphics[width=40pc]{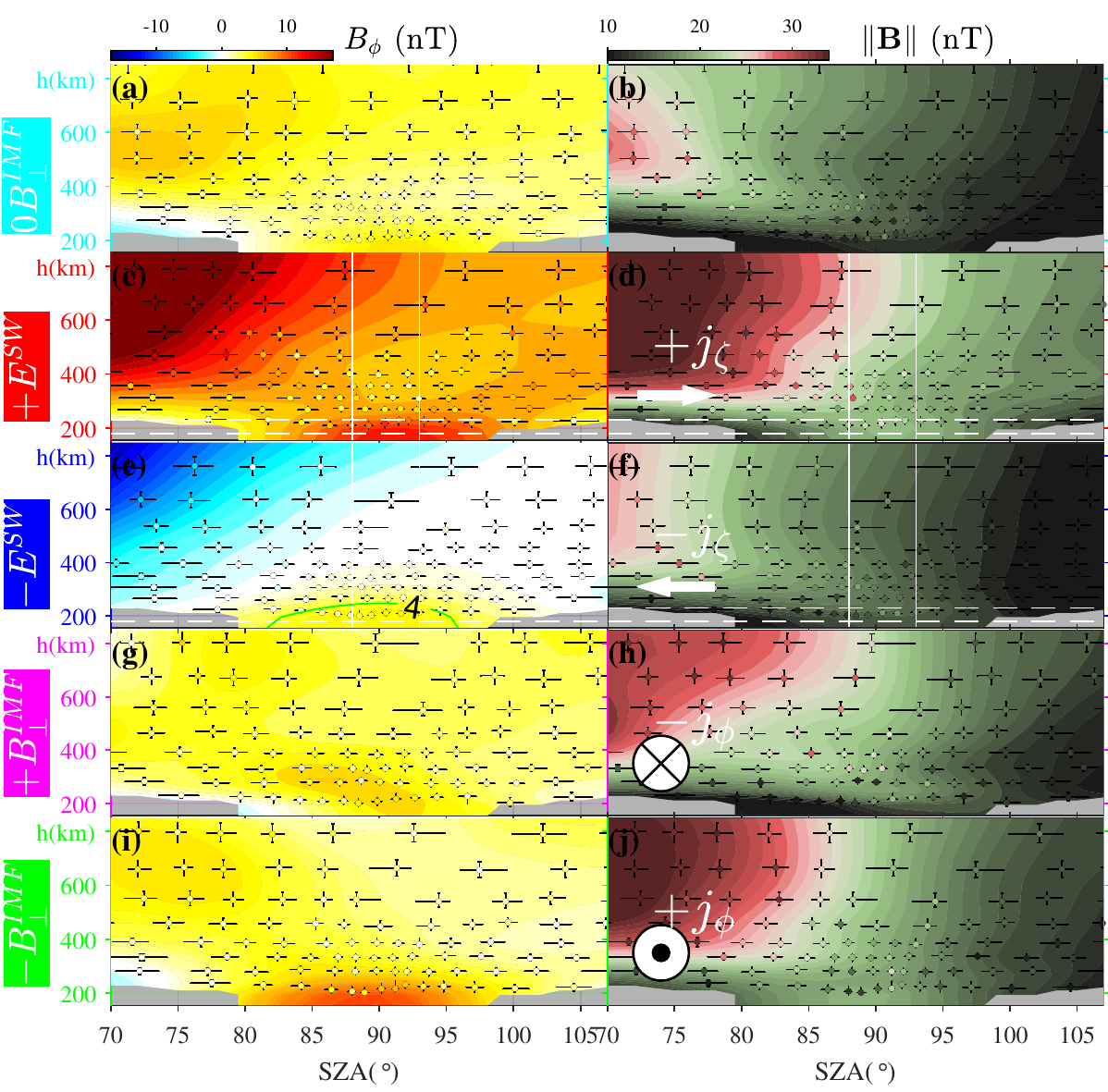}
\caption{The same figures as the Panels in the second and fourth columns in Figure \ref{fig:Bxyz5IMF} but zoomed into SZA-altitude depiction. Shown here is only a small region over the terminator (corresponding to the dashed white line in Figure \ref{fig:Bxyz5IMF}d). In Panels in the second column, the arrow, cross, and dot symbols at SZA=75$^\circ$ denote the orientation of the ionopause current.
}
\label{fig:Bxyz_zoom}
 \end{figure}

 We distribute the $\mathbf{B}^V$ observations under the $0B_{\perp}^{IMF}$ condition, colored in cyan in Figure \ref{fig:IMF_Dist},
 equally into fourteen altitude levels, then into fourteen SZA bins at each altitude level, resulting in 14$\times$14 spatial bins with approximately identical sampling count, e.g., $N_{sample}=$478--479 in all bins.
 In each bin, the median of the position vector ($r_\rho$, $r_\zeta$) of the 478--479 samples is shown as a point in the $\rho$-$\zeta$ plane in Figure \ref{fig:Bxyz5IMF}a and is color-coded with the median of the component $B_\rho$.
 On each point, the black cross is the error bar that represents the upper and lower quartiles of ($r_\rho$, $r_\zeta$) of the 478--479 samples.
 Similarly, $B^V_\phi$ and $B^V_\zeta$ components, and the magnitude $\|\mathbf{\mathbf{B}^V}\|$ are also constructed in
 Figures \ref{fig:Bxyz5IMF}b, \ref{fig:Bxyz5IMF}c, and \ref{fig:Bxyz5IMF}d, respectively. For other four IMF conditions, $\mathbf{\mathbf{B}^V}\left(r_\rho, r_\zeta\right)$ are also constructed and displayed in Figures \ref{fig:Bxyz5IMF}e--h, \ref{fig:Bxyz5IMF}i--l, \ref{fig:Bxyz5IMF}m--p, and \ref{fig:Bxyz5IMF}q--t.
 In addition, to check the details, we zoom and project $B^V_\phi$ and $\|\mathbf B^V\|$, namely,
 Figures \ref{fig:Bxyz5IMF}b, \ref{fig:Bxyz5IMF}d, \ref{fig:Bxyz5IMF}f, \ref{fig:Bxyz5IMF}h, \ref{fig:Bxyz5IMF}j, \ref{fig:Bxyz5IMF}l, \ref{fig:Bxyz5IMF}n, \ref{fig:Bxyz5IMF}p, \ref{fig:Bxyz5IMF}r, and \ref{fig:Bxyz5IMF}t, into the SZA-altitude depiction in Figure \ref{fig:Bxyz_zoom}.

 The IMF dependence of Venusian magnetospheric $\mathbf{B}^V$distribution has been observationally investigated in both cases \citep[e.g.,][]{Zhang2012} and statistical studies \citep[e.g.,][]{Dubinin2014,Zhang2015}, which are mostly based on manually-selected observations and therefore are subject to potential prior expectations. Some studies have also constructed unbiased-statistical $\mathbf{B}^V$ under different SW/IMF conditions \citep{Chai2016,Zhang2010}, using data without any guiding selection.
 These unbiased studies focused mainly on large structures at planetary scales, because they used a popular data averaging approach. The approach divides the VEX samplings spatially uniformly into cubes, and averages data in each cube separately \citep[e.g.,][]{ Du2013, Zhang2010}. The resultant maps have uniform spatial resolutions but non-uniform standard error of the mean (SEM) that is proportional to $1/\sqrt{N_{sample}}$. Here, $N_{sample}$ counts observations in a bin, which is highly spatially non-uniform \citep[Figure~\ref{fig:histo}a, also cf,][]{ Du2013, Chai2016}. In the current study, we map VEX magnetic observations with uniform SEM by averaging the same number of samplings $N_{sample}$ in each bin, which yields a uniform significance with non-uniform spatial resolution, different from the uniformly spatially resolved works \citep[e.g.,][]{Du2013}.

 Our resolution is about 1--1.5$^\circ$ in SZA and up to about 50~km in altitude over 90$^\circ$ SZA at 200~km altitude, at least one order of magnitude higher than the typical resolution of results on the uniform spacing grid \citep[e.g., about 600$\times$600~km in][]{Du2013,Chai2016}.

 \clearpage

 \begin{figure}[h]
	\includegraphics[width=43pc]{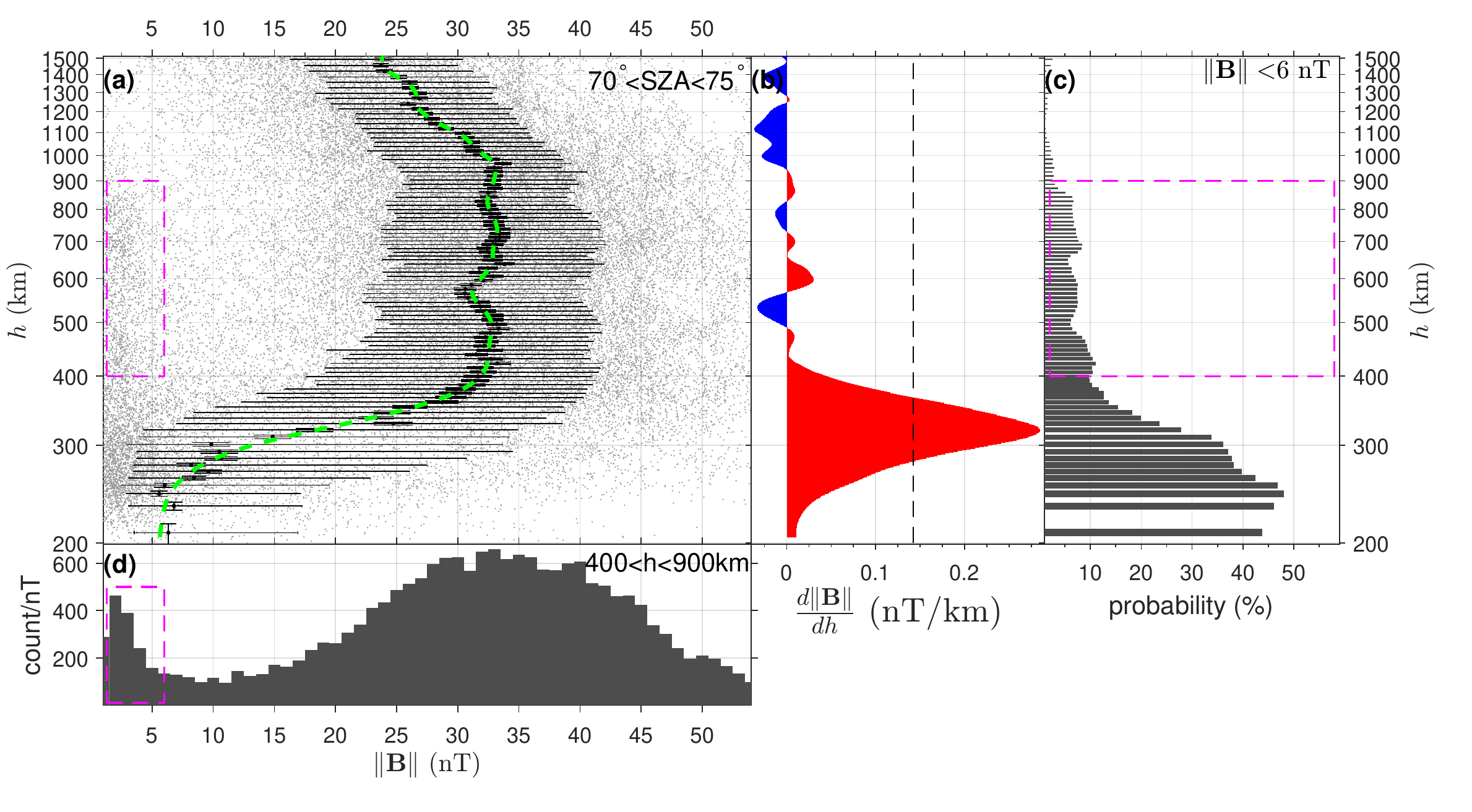}
	\caption{ (a) the induced magnetic field $\|\mathbf{B}\|$, (b) its vertical gradient $\frac{d\|\mathbf{B}\|}{dh}$, and (c) the probability of  $\|\mathbf{B}\|<$5~nT as a function of altitude at $70^\circ<$SZA$<75^\circ$.		(d) histogram of  VEX's observations  between 400 and 900~km altitude at $70^\circ<$SZA$<75^\circ$. In (a), one gray dot denotes one VEX's $\|\mathbf{B}\|$ observation. These samples are divided into altitude bins. Each bin comprises an equal number of samples. In each bin, the quantiles of the altitude and $\|\mathbf{B}\|$ of the samplings are denoted as a black cross in (a) comprising  vertical and horizontal error bars. The green line in (a) illustrates a smoothing spline of the median according to which the gradient in (b) is calculated. In (b), the dashed line denotes the half-maximum of the dominant red peak, which is used to estimate the statistical height and distribution of the ionopause. In Panels (a, c, d), the magenta boxes highlight some special situations discussed in Section~\ref{sec:draping_sym}.}
	\label{fig:IP}
\end{figure}

\begin{figure}[h]
	\centering
	\includegraphics[width=40pc]{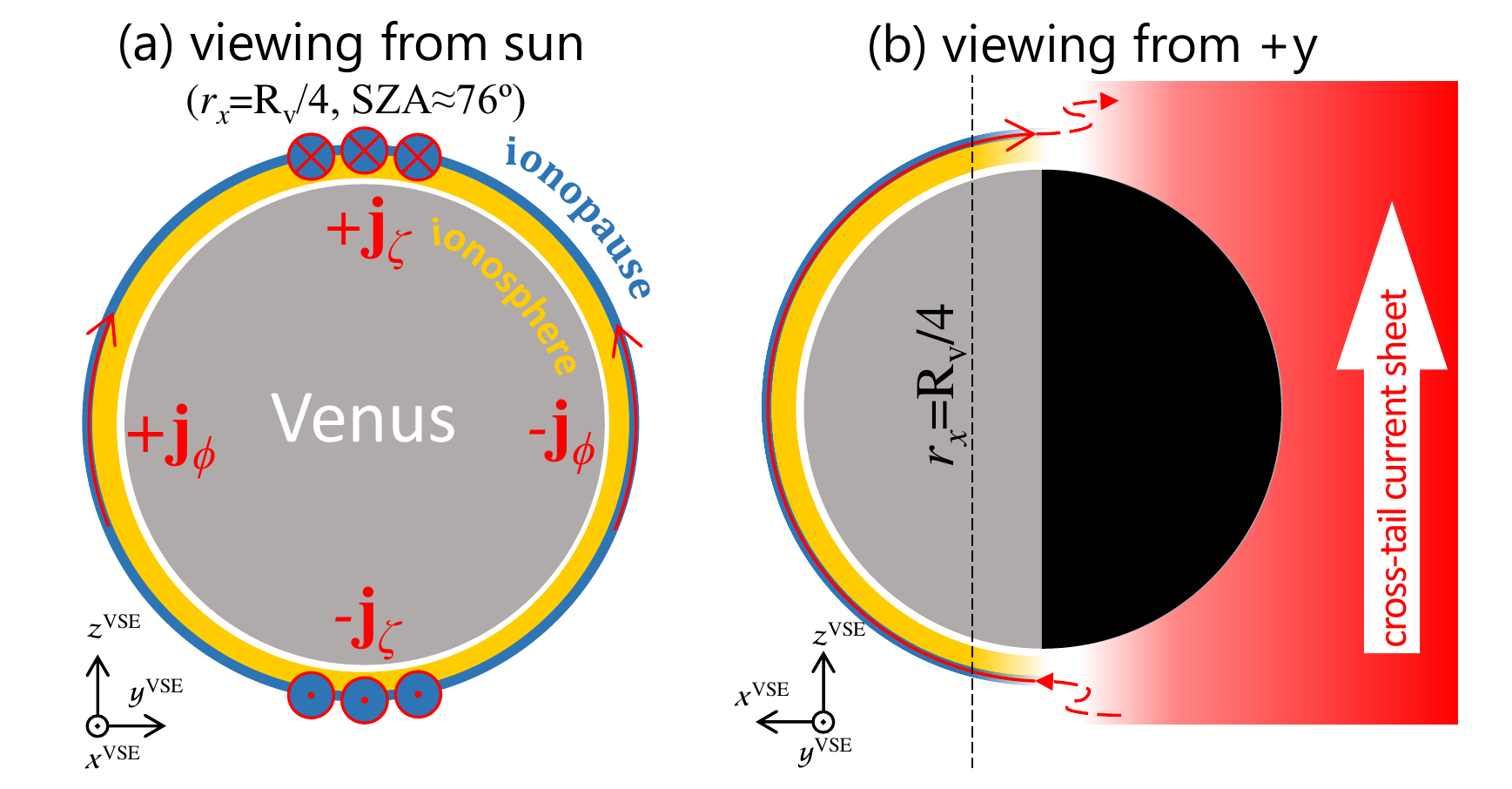}
	\caption{Sketches of the ionopause current in the planes (a) $r_x$=Rv/4, and (b) $r_y$=0 in the VSE coordinate system, viewing from the sun and the +y flank, respectively.
	}
	\label{fig:J_sketch}
\end{figure}

\begin{figure}[h]
	\centering
	\includegraphics[width=40pc]{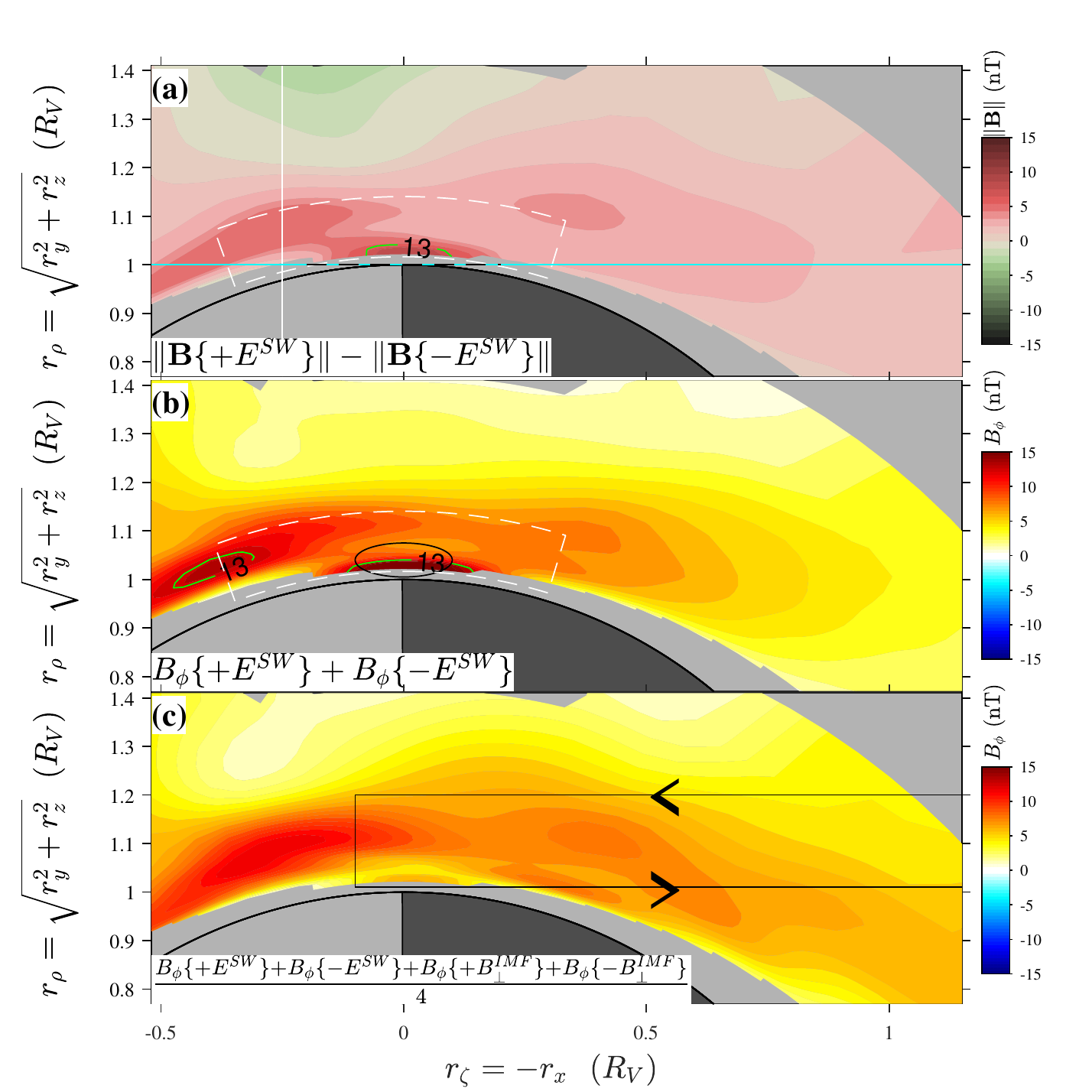}
	\caption{(a) the difference between Figures \ref{fig:Bxyz5IMF}h and \ref{fig:Bxyz5IMF}l, (b) the sum of Figures \ref{fig:Bxyz5IMF}f and \ref{fig:Bxyz5IMF}j, and (c) the average of Figures \ref{fig:Bxyz5IMF}f, \ref{fig:Bxyz5IMF}j, \ref{fig:Bxyz5IMF}n, and \ref{fig:Bxyz5IMF}r. The green contour lines in (a) and (b) highlight the most intense peaks in these Panels. The black arrowed lines in (c) sketch a current proposed in \citet[][]{Chai2016}.
	}
	\label{fig:Symtr_mean}
\end{figure}
 \section{Classical IMF draping configuration at a high spatial resolution} \label{sec:draping}
 The current section illustrates that our mapping approach can resolve structures at various spatial scales, including the planetary-scale IMF draping configuration and the thin ionopause. For convenience, hereafter, we denote the induced $\mathbf{B}^V$ on the in the $\rho$-$\zeta$ plane as $\mathbf{B}\{\bullet\}$=[$B_{\rho}\{\bullet\}$, $B_{\phi}\{\bullet\}$, $B_{\zeta}\{\bullet\}$], where $\bullet\in\{0B_{\perp}^{IMF}$, $\pm B_{\perp}^{IMF}$, $\pm E^{SW}\}$ denotes the IMF conditions.

 \subsection{Symmetries}
 \label{sec:draping_sym}
 In Figure \ref{fig:Bxyz5IMF}, the most notable features in the three magnetic components are three anti-symmetries,
 between $B_\rho\{+B_{\perp}^{IMF}\}$ and $B_\rho\{-B_{\perp}^{IMF}\}$,
 between $B_\phi\{+E^{SW}\}$ and $B_\phi\{-E^{SW}\}$, and
 between $B_\zeta\{+B_{\perp}^{IMF}\}$ and $B_\zeta\{-B_{\perp}^{IMF}\}$,
 illustrated in Panels (m, q), (f, j), and (o, s), respectively.
 The anti-symmetries in $B_\rho$ and $B_\zeta$ reveal the draped IMF: the magnetic field is bending from $\pm\rho$ direction in upstream SW to $\pm\zeta$ direction in the magnetosphere \citep[for a three-dimensional schematic draping pattern, readers are referred to][]{He2016}.

 Although above the three magnetic components exhibit dependence on SW/IMF conditions, $\|\mathbf{B}\|$ exhibits similarities under different SW/IMF conditions. The first similarity is the distinctive day-night difference: on the dayside, e.g., at SZA=70$^\circ$, $\|\mathbf{B}\|$ is stronger than that on the nightside at SZA$>$ 90$^\circ$, and it maximizes vertically at about 400--800~km altitude (see Figures~\ref{fig:Bxyz_zoom}b, \ref{fig:Bxyz_zoom}d, \ref{fig:Bxyz_zoom}f, \ref{fig:Bxyz_zoom}h, and \ref{fig:Bxyz_zoom}j, zoomed versions of Figures~\ref{fig:Bxyz5IMF}d, ~\ref{fig:Bxyz5IMF}h, ~\ref{fig:Bxyz5IMF}l,~\ref{fig:Bxyz5IMF}p, and ~\ref{fig:Bxyz5IMF}t) which corresponds to the magnetic barrier. For example, under $+E^{SW}$ in Figures \ref{fig:Bxyz5IMF}h and \ref{fig:Bxyz_zoom}d, $\|\mathbf{B}\|$ in the magnetic barrier is a few times stronger than the IMF magnitude: $\|\mathbf{B}^{barrier}\|>$ 40~nT vs. $\langle\|\mathbf{B}^{IMF}\|\rangle$= 7.3~nT. Since the $\|\mathbf{B}\|$ in Figures~\ref{fig:Bxyz5IMF} and \ref{fig:Bxyz_zoom} share similarities between different IMF conditions, in Figure~\ref{fig:IP}a we depict $\|\mathbf{B}\|$ as a function of $h$   using all VEX's observations at SZA=70--75$^\circ$ without considering IMF conditions. 
 
 In Figure~\ref{fig:IP}, $\|\mathbf{B}\|$  maximizes at 30--35~nT between 400 and 900~km altitude. In this altitude range, the probability density of $\|\mathbf{B}\|$ exhibits two maxima, at 25--40 and 0--6~nT, as illustrated in Figure~\ref{fig:IP}d. The 25--40-nT maximum corresponds to the magnetic barrier, whereas the 0--6-nT maximum reveals the unmagnetized ionosphere which exists below the ionopause. We display the probability $p$ of $\|\mathbf{B}\|<$6~nT as a function of altitude in Figure~\ref{fig:IP}c to use the  weak $\|\mathbf{B}\|$ as an indicator of the unmagnetized ionosphere. At 400--900~km, for about $p<$10\%, the ionosphere is unmagnetized. (Note that this indicator is imperfect. Otherwise, (1) $p$ should maximize at 100\% at a low altitude and decrease monotonously with $h$, and (2) its vertical gradient should be proportional to the probability density of in-situ occurrences of ionopause.) In Figure~\ref{fig:IP}a, below and above the 400--900~km region,  $\|\mathbf{B}\|$ decreases with decreasing and increasing $h$, respectively. The vertical gradients are signatures of  meridional currents flowing oppositely on the ionopause and IMB, as sketched in Figure 19 in \citet{Baumjohann2010}, which is quantified in Figure~\ref{fig:IP}b. In Figure~\ref{fig:IP}b, the $\|\mathbf{B}\|$ gradient maximizes at $h$=320~km, with a full width at half maximum (FWHM) of about 90~km, extending from 280 to 370~km. Note that the gradient peak at $h$=280--370~km describes the altitude of the maximum probability density of ionopause occurrences, rather than the complete occurring range or thickness of the ionopause. The ionopause at every instant is a thin layer without a vertical extension. The altitude $h$=280--370~km is above the exobase \citep[150--200~km, see Table 2 in][]{Hodges2000}, suggesting that during VEX's operation, the dayside (SZA$<$70--75$^\circ$) bottom ionosphere is unmagnetized for most of the time. The low-ionospheric magnetization reported in VEX observations \citep[e.g.,][]{Dubinin2014,Zhang2012} occurs mostly around the terminator (as detailed below in Section~\ref{sec:Hall_curr_ter}), since most VEX's low-ionospheric magnetic observations are distributed not far from the terminator due to the constraints of VEX's obit as displayed in Figure~\ref{fig:histo} and also pointed out by \citet{Dubinin2014}. At $h<$ 200~km, 75.7\% of the observations are distributed at 85$^\circ\le$SZA$\le$95$^\circ$ (Figure~\ref{fig:histo}b).

 The $\|\mathbf{B}\|$ similarities among Figures~\ref{fig:Bxyz5IMF}d, ~\ref{fig:Bxyz5IMF}h, ~\ref{fig:Bxyz5IMF}l,~\ref{fig:Bxyz5IMF}p, and ~\ref{fig:Bxyz5IMF}t are associated with different magnetic components under different SW/IMF conditions.
 Under $\pm E^{SW}$ (among Figures \ref{fig:Bxyz5IMF}e--g and \ref{fig:Bxyz5IMF}i--k) the gradient is sharper in $B_\phi$ than that in the other two components, whereas under $\pm B_{\perp}^{IMF}$ (among Figures \ref{fig:Bxyz5IMF}m--o and \ref{fig:Bxyz5IMF}q--s) the gradient in $B_\zeta$ is the sharpest. The gradients of these components suggest that the dominant components of the ionopause current in the $\pm E^{SW}$ and $\pm B_{\perp}^{IMF}$ hemispheres are in the directions of $\pm \zeta$- and $\pm \phi$-axis, respectively, as displayed by the arrows, cross, and point in Figures \ref{fig:Bxyz_zoom}d, \ref{fig:Bxyz_zoom}f, \ref{fig:Bxyz_zoom}h, and \ref{fig:Bxyz_zoom}j. Figures \ref{fig:J_sketch}a and \ref{fig:J_sketch}b sketch the topology of the ionopause current in VSE coordinates, in the plane $r_x$=Rv/4 and $r_y$=0, respectively. In the current work, Rv=6052~km denotes Venus radius.
 The currents exhibit symmetry between $\pm B_{\perp}^{IMF}$hemispheres and anti-symmetry between $\pm E^{SW} $ hemispheres. On the dayside, $\pm j_\phi$ in the $\pm B_{\perp}^{IMF}$ hemispheres closes $+j_\zeta$ in the $+E^{SW} $ hemisphere with $-j_\zeta$ in the $-E^{SW}$ hemisphere. Over the poles of the $\pm E^{SW}$ hemispheres, $\pm j_\zeta$ flows across the terminator, and then closes through the SW or through currents on the IMB or in the magnetic barrier. 
 Further, the vertical $B_\phi$ gradient in, e.g., Figures \ref{fig:Bxyz5IMF}f and \ref{fig:Bxyz_zoom}c, allows estimating the ionopause current density: the drop by more than 20~nT from the magnetic barrier maximum to low altitude is associated with a current at an intensity more than 15~A/km under the assumption of sheet-like current distribution.

\subsection{$\pm E^{SW}$ asymmetries} \label{sec:draping_Asym}

Although Section \ref{sec:draping_sym} illustrates the similarity of $\|\mathbf{B}\|$ between the $\pm E^{SW}$ hemispheres, the $\|\mathbf{B}\|$ intensity is stronger in $+E^{SW}$ hemisphere than that in the other. Referring $\|\mathbf{B}\|$=25~nT as a boundary, the magnetic barrier extends to about 85$^\circ$ SZA under the $+E^{SW}$ condition (Figure~\ref{fig:Bxyz_zoom}d) but about 75$^\circ$ SZA under the $-E^{SW}$ (Figure~\ref{fig:Bxyz_zoom}f). To quantify the $\pm E^{SW}$ difference, in Figure~\ref{fig:Symtr_mean}a we display $\|\mathbf{B}\{+E^{SW}\}\|-\|\mathbf{B}\{-E^{SW}\}\|$, namely, the difference between Figures \ref{fig:Bxyz5IMF}h and \ref{fig:Bxyz5IMF}l. 
The main contributing component to the asymmetric $\|\mathbf{B}\|$ is $B_\phi$ which exhibits $\pm E^{SW}$ anti-symmetric polarity (Figures~\ref{fig:Bxyz5IMF}h and \ref{fig:Bxyz5IMF}l) as discussed in Section \ref{sec:draping_sym}.
To quantify the break of anti-symmetry between Figures~\ref{fig:Bxyz5IMF}h and \ref{fig:Bxyz5IMF}l, we display $B_\phi\{+E^{SW}\}+B_\phi\{-E^{SW}\}$ in Figure~\ref{fig:Symtr_mean}b. The $\pm E^{SW}$ difference reflects the asymmetry of the IMF wrapping between the $\pm E^{SW}$ poles, as reported in \citet{Zhang2010}.

In Figure~\ref{fig:Symtr_mean}a, the $\|\mathbf{B}\|$ difference exhibits three peaks, on the dayside in the magnetic barrier, on the nightside in the very near magnetotail, and exactly over the terminator at low altitude (in the ionosphere, see the contour line of 13~nT). The barrier and terminator $\|\mathbf{B}\|$ peaks are associated with peaks in the $B_\phi$ difference in Figure~\ref{fig:Symtr_mean}b (above 13~nT, see the contour lines). The barrier and magnetotail peaks extend into global scales, whereas the terminator one is restricted quite regionally.
The global-scale asymmetries are known as magnetic $\pm E^{SW}$ asymmetry \citep[investigated in both observations and simulations,][]{Saunders1986, Zhang1991, Jarvinen2013}, which also exists at Mars \citep{Vennerstrom2003}. The underlying mechanism is still under debate. Finite Larmor radii effects of pickup ions were suggested to account for the asymmetry \citep{Phillips1987}, but hybrid simulations \citep{Brecht1990} could reproduce the phenomenon without planetary ions, and indicated that the Hall effect contributes significantly to the asymmetry. In addition, global MHD simulations with multifluid treatment at Mars \citep{Najib2011} reproduced the $\pm E^{SW}$ asymmetry, suggesting that the decoupling of separate ion fluids could serve as an alternative explanation.

Compared with the global-scale asymmetries, the regional one over the terminator is relatively less understood and will be discussed below in Section \ref{sec:Hall_curr_ter}.

 \begin{figure}
\centering
\includegraphics[width=35pc]{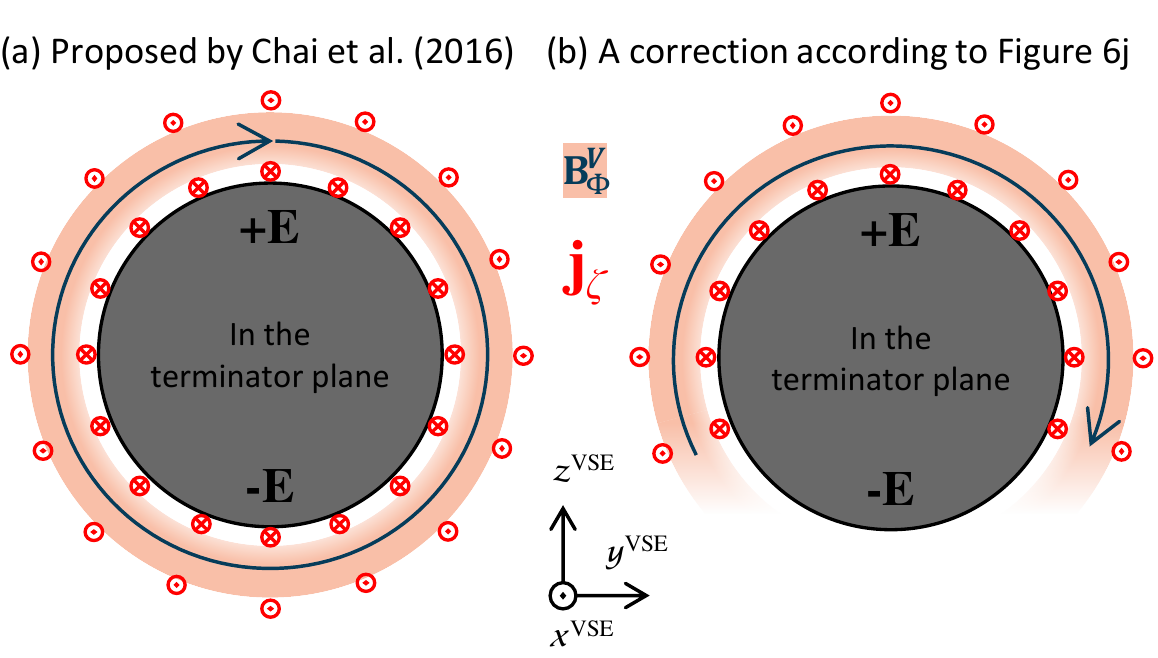}
\caption{ (a) A sketch of the "looping" current and the associated current in the terminator plane according to \citet{Chai2016} and (b) a correction suggested by  Figure~\ref{fig:Bxyz5IMF}j. 
}
\label{fig:sketch_Chai}
 \end{figure}

\section{New lights on the induced field} \label{sec:New_lights}
 The previous section demonstrates the capability of our mapping approach in resolving small-scale strictures.
 In the current section, we investigate two structures that have not been depicted in the classical draping configuration.
 \subsection{ The "looping" magnetic field}
 \label{sec:CB}

 A cylindrical symmetrical preference of positive $B_\phi$ was reported on a cylinder in the magnetotail, which was called the induced global magnetic field "looping" \citep{Chai2016}. The "looping" structure is denoted by the pink region in Figures \ref{fig:Symtr_sketch} and \ref{fig:sketch_Chai}a.
 For a comparison with \citet{Chai2016}, in Figure \ref{fig:Symtr_mean}c we average $B_\phi$ under the SW/IMF conditions $\pm B_{\perp}^{IMF}$ and $\pm E^{SW}$ (Figures \ref{fig:Bxyz5IMF}f, \ref{fig:Bxyz5IMF}j, \ref{fig:Bxyz5IMF}n, and \ref{fig:Bxyz5IMF}r). As a reference, the black box in Figure \ref{fig:Symtr_mean}c sketches the distribution of the $B_\phi$ symmetry according to \citet{Chai2016}.
 In Figure \ref{fig:Symtr_mean}c, the tail-like $B_\phi$ maximum on the nightside is largely consistent with the black box and wrapping slightly towards low latitude. 

 The tail-like $B_\phi$ maximum exhibits a quantitative difference between the different IMF conditions (among Figures \ref{fig:Bxyz5IMF}f, \ref{fig:Bxyz5IMF}j, \ref{fig:Bxyz5IMF}n, and \ref{fig:Bxyz5IMF}r). The $B_\phi$ maximum is strongest under $+E^{SW}$ (Figure \ref{fig:Bxyz5IMF}f) and weaker under $\pm B_{\perp}^{IMF}$ (Figures \ref{fig:Bxyz5IMF}n and \ref{fig:Bxyz5IMF}r). Under $-E^{SW}$ (in the black rectangle in Figure \ref{fig:Bxyz5IMF}j), $B_\phi$ even partially reverses to negative.
 This IMF dependence suggests a cylindrical asymmetry, as sketched in Figure \ref{fig:sketch_Chai}b.
 \citet{Chai2016} explained the symmetry of $+B_\phi$ preference in terms of currents on two cylinders with the identical longitudinal axis pointing towards $\zeta$ and $-\zeta $ \citep[as sketched in Figure 4b in][and here in Figure \ref{fig:sketch_Chai}a by the red symbols]{Chai2016}. The breaking of the symmetry of $+B_\phi$ preference under the $-E^{SW}$ condition reported here suggests that the current system is not cylindrically complete in the near tail. A correction of the "looping" structure and two-cylinder currents is sketched in Figure \ref{fig:sketch_Chai}b.

 This corrected "looping" structure could be explained as the superposition of the classical magnetic draping structure with an additional draping configuration. This superposition theory was proposed for explaining a similar structure on Mars \citep{Dubinin2019}. Similar to Venus, Mars does neither have an internal dipole magnetic field, and its interaction with solar winds produces a similar "looping" structure \citep{Chai2019}. \citet{Dubinin2019} attributed this "looping" structure to an asymmetrical pileup of IMF and an associated additional draping configuration in which magnetic fields bend toward the $-E^{SW}$ direction in the planes normal to the SW velocity. The authors also reproduced this structure in hybrid simulations. The additional draping can produce closed loops over the $-E^{SW}$ hemisphere and weaken the magnetic field and potentially also magnetic reconnection \citep[also see][]{Rong2014,ramstad2020}.

 Another morphological character of this "looping" in Figure \ref{fig:Symtr_mean}c is that it does not overlap with, but is isolated from, the low-ionospheric preference of $B_{\phi}$ over the terminator as detailed in the following subsection. 
 \begin{figure}
\centering
\includegraphics[width=40pc]{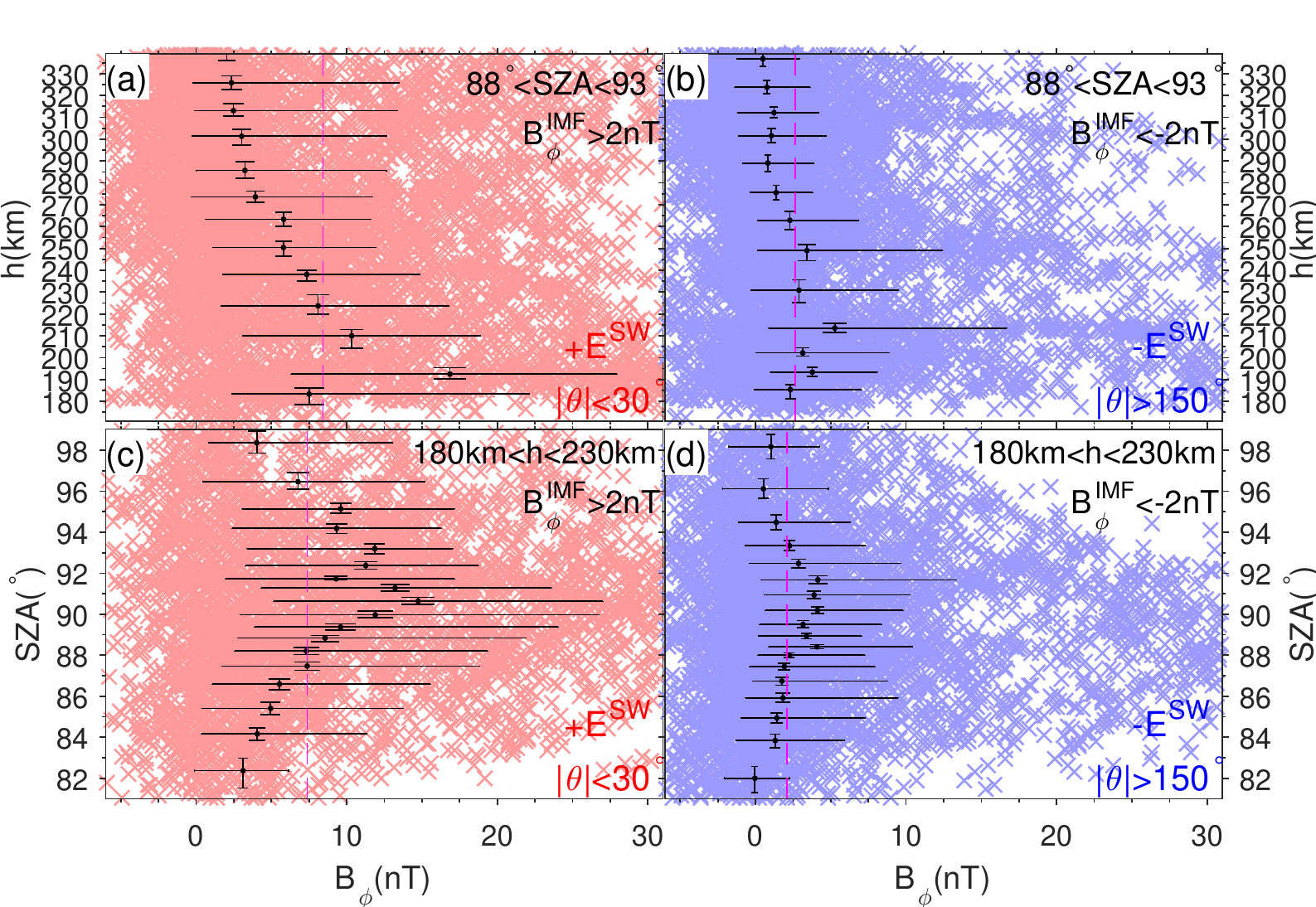}
\caption{(a) the induced magnetic field $B_\phi$ as a function of altitude at $88^\circ<$SZA$<93^\circ$ under the $+E^{SW}$ condition. For this plot, the $B_\phi$ observations are firstly sampled between the vertical white lines in Figure~\ref{fig:Bxyz_zoom}c and divided into altitude bins, each of which comprises an equal number of samples. In each bin, the quantiles of the altitude and $B_\phi$ of the samplings are presented by a black cross comprising a vertical and horizontal error bars. (b) same plot as (a) but for the $-E^{SW}$ condition sampled from Figure~\ref{fig:Bxyz_zoom}e. (c,d) similar plots as (a,b) but for $B_\phi$ as a function of SZA at 180$<h<$230~km altitude (sampled between the horizontal white dashed lines in Figures~\ref{fig:Bxyz_zoom}c and e, respectively). In each panel, one colored cross denotes one VEX sample; the samplings are evenly divided into bins of $h$ or SZA; in each bin, the quartiles of the samplings are denoted as the black bars; and the vertical magenta dashed line denotes the half-maximum (e.g., it occurs at 8.5~nT in (a) because the median of $B_\phi$ maximizes at 17~nT around $h$=190~km).
}
\label{fig:B_localE}
 \end{figure}

 \begin{figure}
 \centering
 \includegraphics[width=25pc]{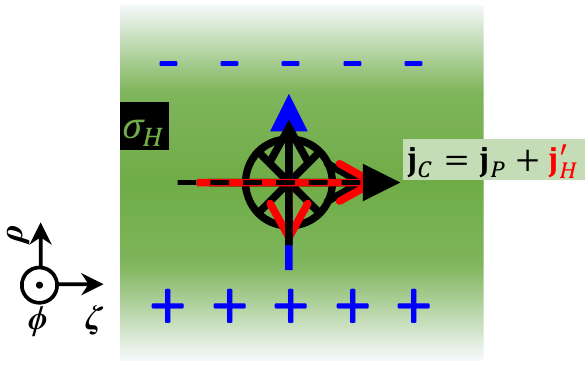}
 \caption{ A sketch of the Cowling channel and Cowling current over the poles of $\pm{E}^{SW}$ hemispheres. The symbols are explained in Figure~\ref{fig:cowling_sketch}a and Appendix \ref{sec:APP}. 
 }
 \label{fig:JcV}
\end{figure}

 \subsection{Ionospheric $\pm E^{SW}$ asymmetry over the terminator} 
 \label{sec:Hall_curr_ter}

 Section~\ref{sec:draping_Asym} mentioned a $\pm E^{SW}$ asymmetry over the terminator (highlighted in the black circle in Figure~\ref{fig:Symtr_mean}b).
 In this small region, the polarity of the induced magnetic component $B_\phi$ is unresponsive to the upstream $B_{\phi}^{IMF}$ polarity and prefers to be positive (see the circle in Figure~\ref{fig:Bxyz5IMF}j and the green isoline in Figure~\ref{fig:Bxyz_zoom}e). Otherwise, in all other regions, the $B_\phi$ polarity is consistent with the IMF polarity of $B_{\phi}^{IMF}$.
 The asymmetry has been investigated using selected cases \citep[e.g.,][]{Dubinin2014, Zhang2015} and was termed as the $\pm E^{SW}$ asymmetrical response of low-ionospheric magnetization to $B_{\perp}^{IMF}$ polarity \citep[e.g.,][]{Dubinin2014}.
 Here, Figures~\ref{fig:Symtr_mean}b, ~\ref{fig:Bxyz_zoom}c and ~\ref{fig:Bxyz_zoom}e present the unbiased statistic at high spatial resolution.

 Not surprisingly, our unbiased statistical results are significantly different from those based on selected cases.
 For example, the maximum $B_{\phi}$ is about 13~nT at +$E^{SW}$ in Figure~\ref{fig:Bxyz_zoom}c and about 4~nT at $-E^{SW}$ in Figure~\ref{fig:Bxyz_zoom}e, which are significantly lower than the averaged magnetic intensity, 45~nT, of the 77 selected events in \citet{Zhang2015}. This significant difference between +$E^{SW}$ and $-E^{SW}$ is also inconsistent with the conclusion in \citet{Zhang2015} that the asymmetry does not show a preference for any particular IMF orientation. Further, our results illustrate that the asymmetry occurs in a very narrow region, about solar zenith angle 85$^\circ<$SZA$<95^\circ$ and altitude $h<$230~km, referring to $B_{\phi}>$4~nT in Figure~\ref{fig:Bxyz_zoom}e. To specify further the extensions in $h$ and SZA, we construct the one-dimensional distribution of $B_{\phi}$ as a function of $h$ and SZA in Figure~\ref{fig:B_localE} under $\pm E^{SW}$ conditions. The vertical dashed line in each panel displays the half-maximum of $B_{\phi}$, referring to which $B_{\phi}$ peaks at about $h$=185--210~km, $h$=185--250~km, SZA=88--95$^\circ$ and SZA=88--93$^\circ$ in Figures~\ref{fig:B_localE}a,~\ref{fig:B_localE}b,~\ref{fig:B_localE}c, and~\ref{fig:B_localE}d, respectively. (Note that $B_{\phi}$ maxima will be smeared out if using broader $h$ or SZA windows. Therefore, the estimation of the $B_{\phi}$ peak width is subject to the sampling window widths.) Among the identifications of these four peaks, the peak identification in Figure~\ref{fig:B_localE}b
  is most vulnerable to disturbances of the half-maximum of $B_{\phi}$. For example, if the half-maximum of $B_{\phi}$ in Figure~\ref{fig:B_localE}b increases by 2~nT, the peak width will shrink from $h$=185--250~km to 200--230~km. Therefore, below we will describe the altitude variations referring mainly to Figure~\ref{fig:B_localE}a.  The regional distribution implicates that simulations to reproduce the structure entail a high spatial resolution, at about ~1$^\circ$ in SZA and 10~km in altitude.
 Note that this SZA range was not included in the simulation in Figure 7b by \citet[][]{Villarreal2015} where the simulation presents the magnetic distribution in the VSO x-y plane at z=1~Rv, as sketched by the yellow line in Figure \ref{fig:Symtr_sketch} and the cyan line in ~\ref{fig:Bxyz5IMF}h and Figures~\ref{fig:Symtr_mean}a.
  \begin{figure}
  \centering
  \includegraphics[width=40pc]{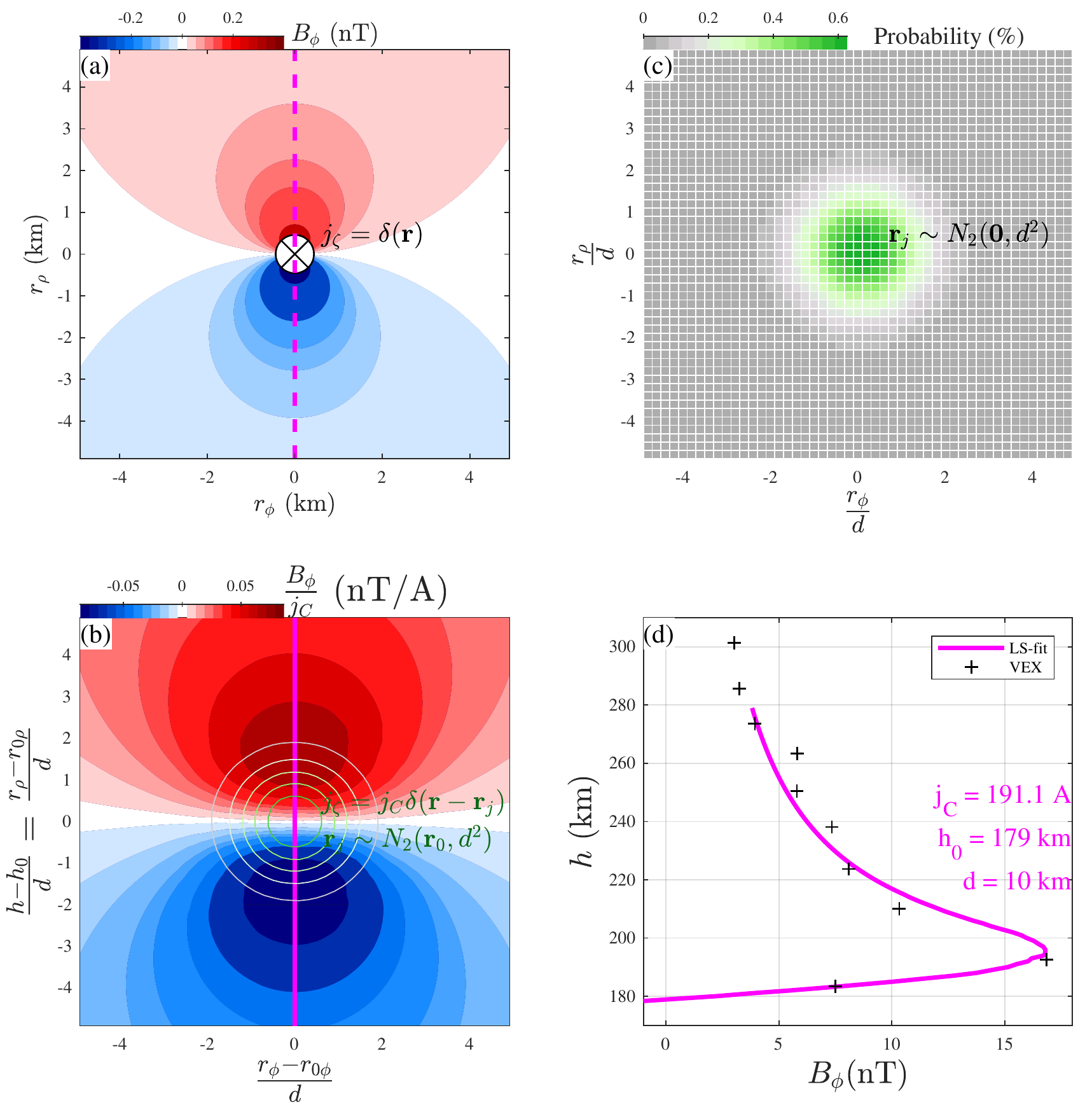}
  \caption{ The expectation of the magnetic component $B_\phi$ induced by (a) $j_\zeta(\mathbf{r})=\delta (\mathbf{r})$ in unit of Ampere,
 namely, a unit line current along $\zeta$ direction centering at the origin of coordinate $\mathbf{r}=\mathbf{0}$, and (b) $j_\zeta(\mathbf{r})=j_C\delta (\mathbf{r}-\mathbf{r}_j)$,
 namely, a unit line current along $\zeta$ direction centering at a random position $\mathbf{r}_j$.
 Here, position vectors $\mathbf{r}$ and $\mathbf{r}_j$ are two-dimensional defined in the $r_{\phi}$-$r_{\rho}$ plane, $\delta$ denotes a Dirac delta function, and $\mathbf{r}_j\sim N_2 (\mathbf{r}_0,d^2)$ denotes a two-dimensional Gaussian random variable with a mean of $\mathbf{r}_0:=[r_{0\phi},r_{0\rho}]$ and variance of $d^2$ as illustrated in (c). (c) The probability distribution of a two-dimensional Gaussian random variable $\mathbf{r}_j\sim N_2 (\mathbf{0},d^2)$. (d) A least-square fit of VEX observations to the model in (b).
In (b) the green and white contours display the same distribution as in (c) but centering at $\mathbf{r}_0$. In (d), the magenta line presents the $B_\phi$ along the magenta line in (b) when $j_C$=191.1~A, $h_0$=179~km, and $d$=10~km, which are fitted using values of the black crosses, namely, the medians displayed as the black dots in Figure~\ref{fig:B_localE}a. Read Section \ref{sec:LS} for details.}
\label{fig:SimuJ}
 \end{figure}

 \subsection{The Cowling channel at Venus}

 Several efforts were made to interpret the asymmetry detailed in Section \ref{sec:Hall_curr_ter}, e.g., in terms of the giant flux rope, magnetic belt, intrinsic field, or induced field, and
  antisunward transports of low-altitude magnetic belts \citep{Zhang2012, Dubinin2014,Villarreal2015}. The potential interpretations were discussed qualitatively in \citet{Dubinin2014} and \citet{Dubinin2014g}, which suggested the most promising explanation is a Cowling current.  As sketched in Figure~\ref{fig:cowling_sketch} and detailed in Appendix A, a Cowling channel entails a particular electromagnetic configuration, including a narrow band of high Hall conductivity $\sigma_H$ elongating perpendicular to an ambient magnetic field $\mathbf{B}_0$, and a primary electric field $\mathbf{E}_{0}$, perpendicular both to the gradient of the $\sigma_H$ band and to $\mathbf{B}_0$.
A Cowling channel might occur over the poles of $\pm{E}^{SW}$ hemispheres with the orientation as sketched in Figure~\ref{fig:JcV}.
 In the daytime Venusian ionosphere, the Hall conductivity $\sigma_H$ maximizes vertically nearby the maximum vertical peak of the ionospheric electron density
 \citep[V2 layer, cf,][]{Dubinin2014, Patzold2007}. The peak is thin vertically and extends broadly in horizontal directions, which is typically under ionopause and therefore is unmagnetized. However, as suggested by Figures~\ref{fig:Bxyz5IMF} and ~\ref{fig:Bxyz_zoom}, the ionopause does not cross the terminator but below about 85$^\circ$ SZA at 200--400~km altitude. Therefore, SW magnetic field $\mathbf{B}_\perp^{IMF}$ might penetrate into the ionosphere over the $\pm{E}^{SW}$ poles \citep{Dubinin2014}. Over the poles, the V2 layer is normal to $\rho$-axis and $\sigma_H$ gradient, while the $\mathbf{B}_\perp^{IMF}$ is parallel to the $\phi$-axis and might serve as $\mathbf{B}_0$. 
 Under this configuration, an antisunward primary electric field $\mathbf{E}_{0\zeta}$ 
might induce a Cowling current along the $\zeta$-axis, $\mathbf{j}_{C\zeta}$.
The magnetic field associated with $\mathbf{j}_{C\zeta}$ might account for the low-ionospheric $\pm E^{SW}$ asymmetry.
However, few issues about the Cowling current are still open. Here we discuss the specific distribution of the current and the primary field $\mathbf{E}_{0\zeta}$.


 \subsubsection{The distribution of the Cowling current} \label{sec:LS}

\citet{Dubinin2014} have not specified how does the Cowling current $\mathbf{j}_{C\zeta}$ distribute. The simplest distribution model is an infinite straight line current with a constant intensity. A line current would induce a magnetic field inversely proportional to the distance to the line, and therefore, the induced magnetic intensity would monotonically decrease with increasing distance. The associated cylindrical component $B_{\phi}$ would also decrease monotonically with increasing altitude above the current, as illustrated by the magenta line in Figure~\ref{fig:SimuJ}a.
 Inconsistent with this monotonic altitude dependence in the current model, the $B_{\phi}$ component of the VEX observation (Figures~\ref{fig:B_localE}a--b) maximizes at about 200~km. To address this inconsistency, we propose a minor correction to the infinite line current model.

We assume that the line current does not occur at a fixed position but at a random position, e.g., $\mathbf{r}_j\sim N_2 (\mathbf{\mathbf{r}_0},d^2)$, a two-dimensional Gaussian random variable with a mean of $\mathbf{r}_0$ and variance of $d^2$. The probability density function of $\mathbf{r}_j$ for the particular case of $\mathbf{\mathbf{r}_0}$:=[$r_{\phi0}$,$r_{\rho0}$]=[0 0] is displayed in Figure~\ref{fig:SimuJ}c.
 A line current with constant intensity $j_C$ at $\mathbf{r}_j$ will induce a distribution of magnetic $B_{\phi}$ as depicted in Figure~\ref{fig:SimuJ}b. Above the current, $B_{\phi}$ exhibits a peak, similar to that in Figure~\ref{fig:B_localE}a. Further, we use values at the black points in Figure~\ref{fig:B_localE}a for a least-square fit to the $B_\phi$ along the magenta line in Figure~\ref{fig:SimuJ}b, resulting in $j_C$=191~A, $d$=10~km, and $r_{\rho0}$=1Rv+179~km. The fitted results, as displayed in Figure~\ref{fig:SimuJ}d, suggest that a current locating at altitude 179$\pm$10km with an intensity of 191~A could account for the $B_{\phi}$ variation in the VEX observation (Figure~\ref{fig:B_localE}a).

 \subsubsection{The primary electric field}
 \citet{Dubinin2014} conjectured that the primary electric field $\mathbf{E}_{0\zeta}$ might arise from downward drifting electrons. In the Cowling channel, electrons are supposed to be decoupled from ions due to their different ratios of the gyro-frequency over the collision frequency ($\kappa_i<<1$ but $\kappa_e>>1$, see Appendix \ref{sec:CC1}).
The downward electron drift is supposed to be at the order of magnitude of 100~m/s and being driven by SW motional electric field $\mathbf{E}^{SW}$ mapping along SW magnetic field lines. However, this mechanism still misses details.
 For example, the downward electron drift is associated with an upward Pedersen current $\mathbf{j}_{P0}$, accumulating potentially polarization charges on the upper and lower ionospheric boundaries. 
 The polarization charges and the induced polarization electric field would superimpose to $\mathbf{E}^{SW}$, resist the downward electron drift and 
 terminate $\mathbf{E}_{0\zeta}$.
 Another detail is that the electron drift mechanism cannot explain why the asymmetry occurs at a very narrow SZA range over exactly the terminator since the Cowling channel and electron drift might occur in a broader SZA range.
 An alternative explanation of $\mathbf{E}_{0\zeta}$ is a horizontal polarization electric field.
 Over the terminator, maximizes the solar radiation gradient in $\zeta$ direction, and therefore maximize the gradients of the ionospheric density $\|\nabla_\zeta {n_e}\|$ and conductivity $\|\nabla_\zeta {\sigma_P}\|$. It is also over the terminator in the $\pm E^{SW}$ hemisphere where the ionopause current merges the cross-tail current. By preventing a divergence of currents, polarization charges and electric field $\mathbf{E}_\zeta$ might be created and serve as
 the primary electric field $\mathbf{E}_{0\zeta}$ in the Cowling channel.
 Note that the polarization mechanism might be equivalent to the electron drift mechanism but might describe the equilibrium from a different perspective.
 Evaluating mechanisms and quantitatively explaining the equilibrium entail simulations at a spatial resolution higher than the order of magnitude 10~km due to the narrow vertical extension.

 \section{Summary}
 The current work uses all VEX magnetic observations to map the near-Venus induced magnetic field under different IMF conditions, at a high spatial resolution through an unbiased statistical method. Using VEX SW observations, we predict the IMF for VEX's magnetospheric transit and evaluate the prediction. For the mapping, we define the Solar-Venus-cylindrical (SVC) coordinate system. Similar to the VSE system, SVC also makes use of the cylindrically symmetrical response of the induced magnetic field to IMF, but is superior to VSE because the determination of SVC axes is not subject to the error of the IMF estimation and SVC can deal with the near-zero IMF condition.
 Our mapping resolution is spatially not uniform, maximizing in the ionosphere where VEX collects most observations. The high resolution enables depicting the small-scale structures, such as the ionopause and the associated, as well as the planetary-scale structures, such as the classical draping configuration and the magnetic pileup region. At 70--75$^\circ$ solar zenith angle (SZA), the ionopause occurs with the highest probability density between 280 and 370~km altitude, above the expected exobase (150--200~km),  suggesting that the bottom ionosphere at SZA$<$75$^\circ$ is unmagnetized for most of the time during VEX's operation.  
 Our mapping reveals that the magnetic pileup region extends to about 85$^\circ$ SZA over the $+E^{SW}$ pole but about 75$^\circ$ SZA over the $-E^{SW}$ pole. In addition, our mapping resolves the $\pm E^{SW}$ asymmetry in the low-ionospheric magnetization and the planetary-scale "looping" magnetic field. The "looping" structure is cylindrically asymmetrical: much stronger under $+E^{SW}$ than under $-E^{SW}$. Under $-E^{SW}$, the "looping" breaks, which can be attributed to the presence of an additional IMF draping configuration occurring in the planes perpendicular to the SW velocity. The "looping" occurs in the magnetotail, separated from the low-ionospheric $\pm E^{SW}$ asymmetry occurring in the narrow region, at about 88--95$^\circ$ SZA and 185--220~km altitude. Our least-square fit of VEX observations to a line current model at a Gaussian random position suggests that the altitude variation of the low-ionospheric asymmetry is attributable to an antisunward line current with an intensity of 191.1~A at 179$\pm$10~km altitude. We explain this line current as in terms of a Cowling channel.  

 Our results also implicate that to reproduce the Venusian low-ionospheric magnetization through simulations entails an altitude resolution of about 10~km. As artificial satellites orbit extraterrestrial planets mostly with high orbital eccentricities, their observations are typically highly spatially uneven distributed. Our mapping method  exploits these unevenly distributed datasets to resolve structures on various scales, with higher resolutions in regions with denser observations. Further works could implement the method on observations at Mars and Titan where similar low-ionospheric magnetization and the planetary-scale "looping" structures are expected or observed.

 \begin{acknowledgments}
VEX magnetic data are available in the ESA's Planetary Science Archive. MH and ED acknowledge the supports from the Deutsche Forschungsgemeinschaft through grants HE6915/1-1 and TE664/4-1, respectively.
 \end{acknowledgments}

 \clearpage
 \appendix
 \renewcommand\thefigure{\thesection.\arabic{figure}}
 \section{Cowling channel} \label{sec:APP}
 \subsection{Atmospheric electrical conductivity and its altitude dependence} \label{sec:CC1}
 In magnetized plasma, Lorenz force drives charged particles drifting gyratorily and periodically perpendicular to the ambient magnetic field ${{\mathbf{B}}_0}$. As a result, the particles are bound to magnetic field lines. The gyratory drift might be interrupted by collisions with other particles, either charged or neutral. In an extreme collisional case, the gyratory is relatively negligible, and particles are freed from the magnetic field and move freely at the bulk velocity of the ambient particles. A measure of the relative importance between the gyratory drift and collision is the ratio of the gyro-frequency to collision frequency $\kappa=\frac{\Omega}{\nu}$. The ratio $\kappa $ quantifies the expectation of the number of gyratory drifting circles that one particle could achieve in-between twice collisions with other particles. The collision frequency is proportional to particle densities. Therefore, in planetary atmospheres, the collision frequency decreases exponentially with increasing altitude as the neutral density is decreasing, whereas the gyro frequency is relatively less dependent on altitude.
 Accordingly, $\kappa $ is overall increasing with altitude. At low altitude, collisions are superior to gyratory drifts ($\kappa <<1$) whereas at high altitude, the collisions are rare and gyratory drifts are more important ($\kappa >>1$). In-between the low and high altitudes, there is a transitional altitude $h^{\kappa \approx1}$, where $\kappa \approx 1$.
However, the vertical increasing rate $\frac{d\kappa }{dh}$ is different for different species, steeper for electrons than for ions $\frac{d{{\kappa}_e}}{dh}>\frac{d{{\kappa }_i}}{dh}$, yielding higher transitional altitudes for ions than for electrons $h_i^{\kappa \approx1}>h_e^{\kappa \approx1}$. Here, $i$ and $e$ in the subscripts denote ions and electrons, respectively.

 As a result, the atmospheric electrical conductivity and its isotropy are altitude-dependent.
 Above $h_i^{\kappa \approx1}$ when plasma is collision-less (${{\kappa }_i}>>1$, ${{\kappa }_e}>>1$), in response to an applied electric field ${\mathbf{E}_0}$, both ions and electrons drift towards the direction of ${\mathbf{E}_0}\times {{\mathbf{B}}_0}$ on the macroscopic scale, and no net current is generated.
 Under $h_e^{\kappa \approx1}$ in the collisional case (${{\kappa }_i}<<1$, ${{\kappa }_e}<<1$), the applied field ${\mathbf{E}_0}$ drives ions and electrons flowing along ${\mathbf{E}_0}$ oppositely, generating a current ${\mathbf{j}_P}={\sigma_P}{\mathbf{E}_0}$ parallel to ${\mathbf{E}_0}$, independent of ${{\mathbf{B}}_0}$. The current ${\mathbf{j}_P}$ is known as Pedersen current and the coefficient ${{\sigma}_P}$ is known as Pedersen conductivity.
 Between $h_i^{\kappa \approx1}$ and $h_e^{\kappa \approx1}$ in the intermediate case ( $\kappa_i<<1$, $\kappa_e>>1$), electrons are already bound to ${{\mathbf{B}}_0}$, whereas ions could still essentially move with ambient winds. The applied field ${\mathbf{E}_0}$ decouples electrons from ions: electrons drift in the direction ${\mathbf{E}_0}\times {{\mathbf{B}}_0}$, whereas the ions prefer following towards ${\mathbf{E}_0}$. The caused current comprises a component ${\mathbf{j}_P}$ following along ${\mathbf{E}_0}$ and a component $\| {\mathbf{j}_H} \|={\sigma_H}\| {\mathbf{E}_0} \|$ parallel to ${\mathbf{E}_0}\times {{\mathbf{B}}_0}$. The current ${\mathbf{j}_H}$ is known as Hall current and the coefficient ${\sigma_H}$ is known as Hall conductivity.

 For references, the range ($h_i^{\kappa \approx 1}$, $h_e^{\kappa \approx 1}$) corresponds approximately to (90km, 150km) in the terrestrial equatorial ionosphere \citep[cf Figure 2.5,][]{kelley2009earth} and (130km, 270km) at Venus \citep[cf,][]{Dubinin2014}. 

 \setcounter{figure}{0}
 \begin{figure}
\centering
\includegraphics[width=40pc]{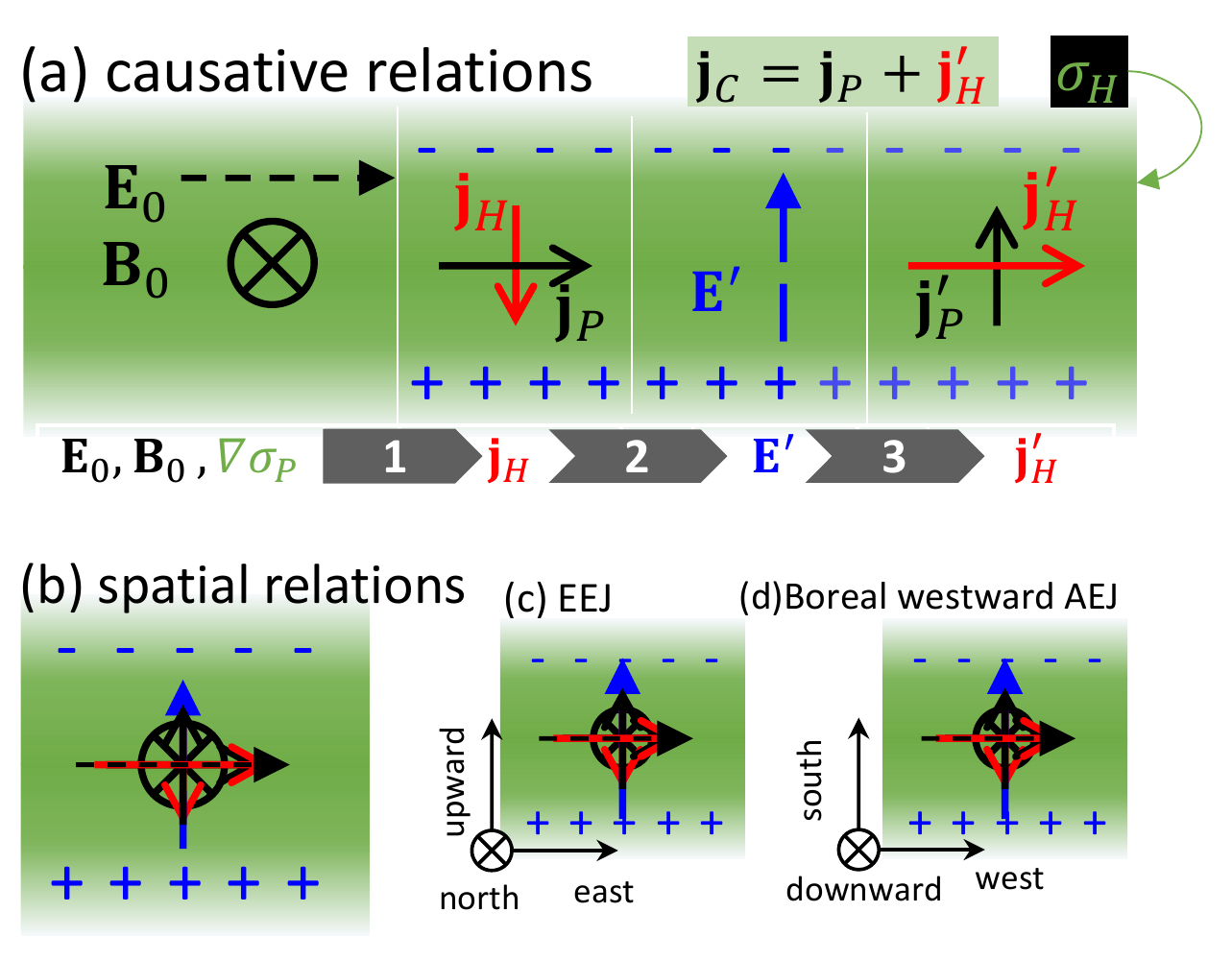}
\caption{ (a) an explanation of a Cowling channel. Note that in Panel (a), the symbols from left to right represent the causative relations rather than the spatial distribution which is sketched in (b). (c, d) orientations of the Cowling channel in equatorial electrojet and boreal westward auroral electrojet.}
\label{fig:cowling_sketch}
 \end{figure}

 \subsection{Cowling channel: a model and two instances at Earth} \label{sec:CC2}
 In a particular configuration of conductivity and magnetic field, the electric current in the direction of ${\mathbf{E}_0}$ might be much intenser than ${\mathbf{j}_P}$ due to a superposition of ${\mathbf{j}_H}$. 
 The geometry of configuration is characterized by a band of ${\sigma_P}$ elongating infinitely in one dimension but restricted in a perpendicular direction. Figure~\ref{fig:cowling_sketch}a sketches the configuration and the causative relations of the superposition through three steps. 
 In the first step, a primary electric field ${\mathbf{E}_0}$, parallel to the elongating direction and within the ambient magnetic field ${\mathbf{B}_0}$ that is normal to the plane spanned by $\nabla {\sigma_P}$ and $ {\mathbf{E}_0}$, generates Hall and Pedersen current ${\mathbf{j}_H}$ and ${\mathbf{j}_P}$. Secondly, on the boundary where $\|\nabla {\sigma_P}\|>$0, polarization charges and electric field $\mathbf{E}'$ are created by preventing a divergence of ${\mathbf{j}_H}$.
 In the last step, the secondary field $\mathbf{E}'$ further generates secondary Hall and Pedersen currents $\mathbf{j}'_H$ and $\mathbf{j}'_P$. $\mathbf{j}'_H$ flows along ${\mathbf{j}_P}$, and therefore, the total current parallel to ${\mathbf{E}_0}$ equals to ${\mathbf{j}_C}$=${\mathbf{j}_P}$+$\mathbf{j}'_H$, intenser than ${\mathbf{j}_P}$. The reinforcement of the current in ${\mathbf{E}_0}$ direction is called Cowling effect \citep[e.g.,][]{Yoshikawa2013}, the reinforced current ${\mathbf{j}_C}$ is known as Cowling current \citep[e.g.,][]{Tang2011}, ${\sigma_C}:=\|{\mathbf{j}_C}/{\mathbf{E}_0} \|$ is Cowling conductivity, and the whole current system is called the Cowling channel. Note that the horizontal separations of elements shown in Figure~\ref{fig:cowling_sketch}a represent only the causative relations in the explanation but not the spatial distribution. In space, the elements coexist under an electrodynamic equilibrium, as sketched in Figure~\ref{fig:cowling_sketch}b. The three steps illustrated in Figure~\ref{fig:cowling_sketch}a are not a unique explanation for the equilibrium. In the end, only the equilibrium matters.

For detailed discussions, readers are referred to the literature in the context of two instances of Cowling channels in Earth's ionospheric E-layer, over the equator and auroral zone, respectively \citep[e.g.,][and references therein]{kelley2009earth,Fujii2011}. The corresponding Cowling currents in these two instances are well known as the equatorial electrojet \citep[EEJ,][and references thereafter]{Forbes1981} and auroral electrojet \citep[AEJ,][and references thereafter]{Bostroem1964}. The orientation of these two Cowling channels are sketched in Figures~\ref{fig:cowling_sketch}c and~\ref{fig:cowling_sketch}d. In both cases, the Cowling channels are zonally-elongated and the geomagnetic field serves as $\mathbf{B}_0$. In the case of EEJ, $\mathbf{B}_0$ is northward, the gradient $\nabla \sigma _P$ is vertical, and the eastward component of the electric field induced by the tidal wind serves as the primary electric field $\mathbf{E}_0$. For the boreal westward AEJ, $\mathbf{B}_0$ is downward, $\nabla \sigma _P$ is in the magnetic meridional direction, and the zonal electric field mapped from the magnetosphere serves as ${\mathbf{E}_0}$.


 \bibliography{Venus}{}
 \bibliographystyle{aasjournal}



\end{document}